\documentclass[pre,twocolumn,superscriptaddress]{revtex4}
\usepackage{graphicx}
\usepackage{float}
\usepackage{flafter}
\usepackage{float}
\usepackage{epsfig}
\usepackage{amsmath, amsthm, amssymb}

\def\gapx{\ \lower 2pt \hbox{$\buildrel>\over{\scriptstyle{\sim}}$}}
\def\lapx{\ \lower 2pt \hbox{$\buildrel<\over{\scriptstyle{\sim}}$}}
\def\lap{\ \lower 2pt \hbox{$\buildrel<\over{\scriptstyle{\sim}}$}}
\def\2d{$2\Delta$}
\def\3he{$^3$He}
\def\4he{$^4$He}

\def\a1{$a_{1}$}
\def\a2{$\overline\alpha_{2}$}

\def\Am3{\AA$^{-3}$}
\def\ax2{$a_{2}$}

\def\d2o{D$_2$O}

\def\h2o{H$_2$O}

\def\Ke3{$\langle{K_3}\rangle$}
\def\k2{$\langle{[k_{\alpha}]^2}\rangle$}

\def\rqw{$R(\textbf{Q},\omega)$}
\def\rw{$R(\omega)$}

\def\sqwe{$S(\textbf{Q},\omega~=0)$}
\def\sqw{$S(\textbf{Q},E)$}

\def\s1{$S_1(\textbf{Q},E)$}

\def\sqw{$S(\textbf{Q},\omega)\;$}

\def\u0{$(u=0)$}

\def\x0{$x=0$}

\def\td{$T_D$}

\def\tse{$T\simeq 100$}

\def\t150{$T\sim150$}

\def\tse2{$T\simeq 250$}

\def\w2l{$\omega^2=\omega _L^2=\phi_L/M$}

\def\m20{$M=20$}
\def\d2o{$D_{2}O$}

\def\a2{$\AA^2$}

\def\cn2{$\alpha$}

\def\rs{$\langle r^2 \rangle$}
\def\rsexp{$\langle r^2 \rangle_{exp}$}
\def\rsslope{$\langle r^2 \rangle_{slope}$}
\def\oqw{$O(\textbf{Q},\omega)$}
\def\oqwe{$O(\textbf{Q},\omega = 0)$}
\def\td{$T_D$}
\def\iqt{$I(\textbf{Q},t)$}
\def\ii{$I_{\infty}$}

\begin{document}
\title{Intrinsic Mean Square Displacements in Proteins}
\author{Derya Vural}                                                                                          
\affiliation{Department of Physics and Astronomy, University of Delaware, Newark, Delaware 19716-2570, 
USA}
\author{Henry R. Glyde}                                                                                         
\affiliation{Department of Physics and Astronomy, University of Delaware, Newark, Delaware 19716-2570, 
USA}     

\date{\today}

\begin{abstract}
The thermal mean square displacement (MSD) of hydrogen in proteins and its associated hydration water is 
measured by neutron scattering experiments and used an indicator of protein function. The observed MSD
as currently determined depends on the energy resolution width of the neutron scattering instrument 
employed. We propose a method for obtaining the intrinsic MSD of H in the proteins, one that is 
independent of the instrument resolution width. The intrinsic MSD is defined as the infinite time value of 
\rs~ that appears in the Debye-Waller factor. The method consists of fitting a model to the resolution 
broadened elastic incoherent structure factor or to the resolution dependent MSD. 
The model contains the intrinsic MSD, the instrument resolution width and a rate constant characterizing 
the motions of H in the protein. 
The method is illustrated by obtaining the intrinsic MSD \rs~ of heparan sulphate (HS-0.4), Ribonuclease A 
and Staphysloccal Nuclase (SNase) from data in the literature.
\end{abstract}
\maketitle

\section{Introduction}
Following pioneering experiments and subsequent developments, the mean square displacement (MSD) of 
hydrogen in proteins can now be readily observed in neutron scattering 
experiments\cite{Doster:89,Doster:90,Zaccai:00,Dellerue:00a,Gabel:02,Sakai:09,Doster:10,Paciaroni:02,Chen:
08b,Magazu:10,Magazu:11}.  Specifically, the global average MSD of H throughout the protein is typically 
determined from the elastic component of the incoherent dynamic structure factor (DSF). In proteins at low 
temperature, the MSD is small. As temperature is increased the MSD increases and often goes through a 
marked increase at a specific temperature or temperatures, \td, denoted the dynamical 
transition.\cite{Doster:89,Doster:90,Zaccai:00,Bicout:01,Roh:05,Khodadadi:08,Hong:11,Becker:04} The onset 
of large values of MSD are associated with the onset of function in 
proteins.\cite{Frauenfelder:91,Ferrand:93,Zaccai:00,Snow:02,Pieper:08,Biehl:08,Zhu:11} Essentially, the 
large amplitude MSD enables contact between different parts of the protein which promotes chemical 
activity, function and possible folding. A large MSD is used as an indicator that function in a protein is 
possible.

In most current methods of data analysis, the MSD and \td~ extracted from experiment depend on the energy 
resolution of the neutron scattering instrument employed. Different MSD \rsexp~ are extracted from data 
observed on different instruments. The higher the resolution, the larger that apparent \rsexp~ observed 
and the lower the apparent \td~ observed. A high energy resolution instrument is needed to observe all the 
motions that contribute to the MSD, including the slow, long time motions.  Indeed, measurements in the 
same protein at the same hydration level on different instruments have been made explicitly to demonstrate 
the dependence of \rsexp~ on the instrument resolution.\cite{Wood:08,Jasnin:10,Nakagawa:10} 

\begin{figure}[ht!]
\hspace{-0.1cm}
\vspace{-0.5cm}
\includegraphics[scale=0.29,angle=0]{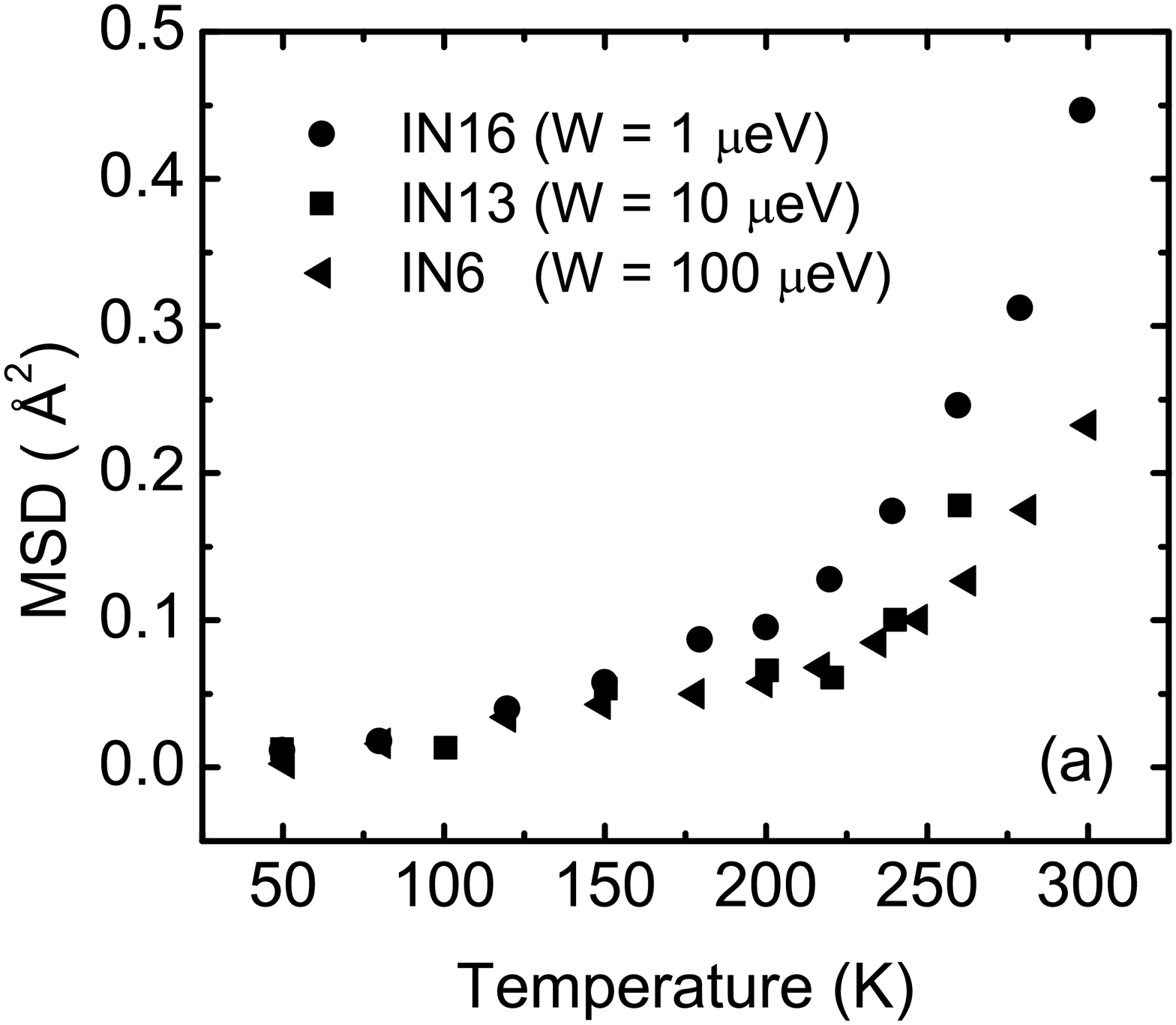}\\
\hspace{0.05cm}
\vspace{-0.5cm}
\includegraphics[scale=0.29,angle=0]{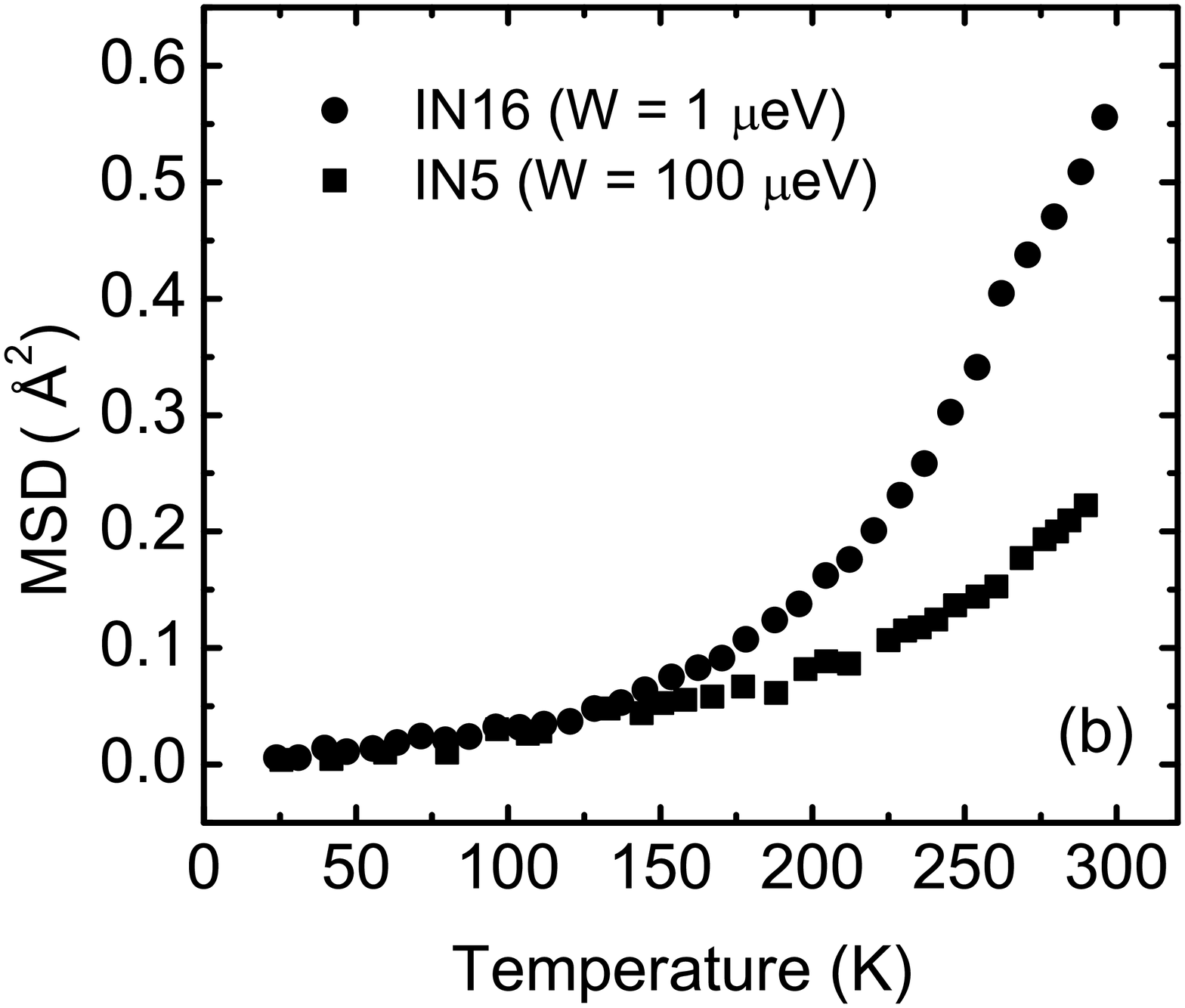}\\
\hspace{0.05cm}
\vspace{-0.5cm}
\includegraphics[scale=0.29,angle=0]{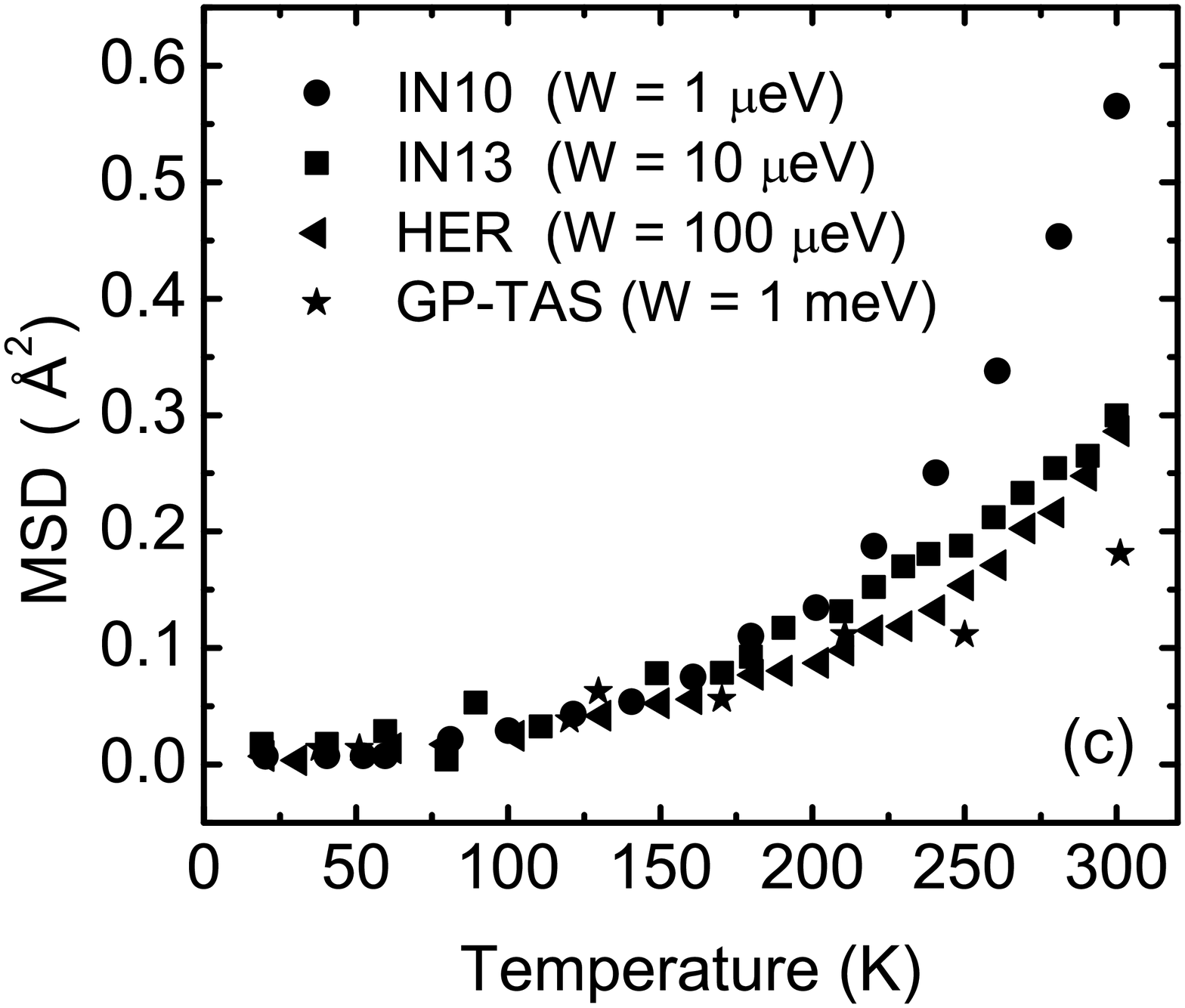}
\vspace{0.8cm}
\caption{MSD of $H$ in proteins observed on neutron scattering instruments having different energy 
resolution  widths, W: (a) in hydrated heparan sulphate (HS-0.4) observed by Jasnin et. al\cite{Jasnin:10} 
on IN16 ($W = 1~\mu$eV), IN13 ($W = 10~\mu$eV) and IN6 ($W = 100~\mu$eV), (b) in hydrated Ribonuclease A 
observed 
by Wood et. al\cite{Wood:08} on IN16 ($W = 1~\mu$eV) and IN5 ($W = 100~\mu$eV), and (c) in hydrated 
Staphysloccal Nuclase (SNase) observed by Nakagawa et. al\cite{Nakagawa:10} on IN10 ($W = 1~\mu$eV), IN13 
($W = 10~\mu$eV), HER ($W = 100~\mu$eV) and GP-TAS ($W = 1~$meV).}
\label{f1}
\end{figure}

Specifically, the observed \rsexp~ is typically obtained from the observed, resolution broadened elastic 
incoherent DSF, \oqwe, as a function of wave vector transfer, $ \textbf{Q} $, as,
\begin{equation} \label{e1}
\langle r^2\rangle _{exp} = -3\frac{d \ln O_{exp}(\textbf{Q},\omega =0)}{dQ^2}.
\end{equation}
Examples of this \rsexp~ obtained on different instruments which displays the dependence of \rsexp~ in the 
instrument energy resolution width are shown in Fig. \ref{f1}. 

The goal of the present paper is to extract the intrinsic MSD, \rs, from the neutron scattering data. This 
is the intrinsic MSD independent of the instrument resolution width, $W$. The intrinsic MSD is defined 
here as the {$\langle r^2 (t = \infty)\rangle$ = \rs~ value of the MSD that appears in the full 
Debye-Waller factor. The intrinsic MSD includes motions up to t = $\infty$. It is the MSD that would be 
observed on an instrument with zero energy resolution width, $W \rightarrow 0$. 

To obtain \rs~ from data, a simple model of the observed incoherent DSF, \oqw, which includes the \rs, the 
resolution width W and a simple description of the motional processes with a single relaxation parameter, 
$\lambda$, is developed. The simple model for the normalized, resolution broadened \oqwe~ is,
\begin{equation} \label{e2}
 O_N(\textbf{Q},\omega = 0) = I_{\infty} + (1-I_{\infty})\frac{W}{(W+\lambda)} + A
\end{equation}
where $I_{\infty}$ is the Debye-Waller factor,
\begin{equation} \label{e3}
I_{\infty} = \exp (-\frac{1}{3} Q^2\langle r^2\rangle).
\end{equation}
and $A$ is an additive constant included if the data includes a constant.
Eq.~(\ref{e3}) is the definition of the intrinsic $\langle r^2\rangle$.
The model is fitted to the observed \oqwe~ for a given W with \rs~ and $\lambda$ treated as free 
parameters to be determined by best fit. An intrinsic \rs~ independent of W is obtained from fits to 
existing data in this way.

As in experiment, Eq. (1), a MSD \rsslope~ obtained from the slope of the model \oqwe~ given by,

\begin{eqnarray} \label{e4}
\langle r^2\rangle _{slope} &=& -3\frac{d \ln O_N(\textbf{Q},\omega =0)}{dQ^2}      \nonumber \\
   &=& \langle r^2\rangle / [1+\frac{W}{I_{\infty} \lambda}].
\end{eqnarray}
can be introduced. As with \rsexp, this \rsslope~ depends on the instrument resolution width W, 
specifically on the ratio $W/\lambda$. The expression for \rsslope~ corresponds to the \rsexp~ and can 
also be fitted to the observed values of \rsexp~ such as shown in Fig.~\ref{f1} to obtain the intrinsic 
\rs. The intrinsic \rs~ is always greater than \rsslope. The intrinsic \td~ is defined as the temperature 
at which the intrinsic \rs~ shows a marked increase with increasing temperature. There can be more than 
one \td\cite{Roh:05,Hong:11}.

In the following section, we develop the model incoherent DSF. In section 3, the model is fitted to 
observed values of the elastic incoherent DSF found in the literature to obtain the intrinsic \rs~ and 
\td~ in three proteins. The intrinsic \rs~ and other parameters are discussed in section 4, where 
suggestions for making the present simple, illustrative model more sophisticated are made. 

\section{Formulation of the Model}

\subsection{Dynamical Structure Factor}

Neutrons incident on proteins interact and scatter predominantly from the hydrogen nuclei in the proteins 
and in the  associated hydration water. The hydrogen nucleus, the proton, has a large, incoherent 
scattering cross-section for  neutrons, 82 barns, which dominates all others. Neglecting the scattering 
from the electrons and other nuclei, the observed scattering intensity is proportional to the incoherent 
dynamical structure factor (DSF), $S_{inc}(\textbf{Q},\omega)$ of the H,
\begin{equation} \label{e5}
 S_{inc}(\textbf{Q},\omega) = \frac{1}{2\pi} \int _{-\infty}^{\infty} dt \exp(i\omega 
t)I_{inc}(\textbf{Q},t),
\end{equation}
where 
\begin{equation} \label{e6}
I_{inc}(\textbf{Q},t)=\frac{1}{N} \sum_{i=1}^{N}\langle \exp(-i\textbf{Q}\cdot\textbf{r}_{i}(t)) 
\exp(i\textbf{Q}\cdot\textbf{r}_{i}(0)) \rangle
\end{equation}
is the intermediate incoherent DSF. In Eq.~(\ref{e5}), $\hbar \textbf{Q}$ and $\hbar\omega$ are the 
momentum and energy, respectively, transferred from the neutron to the protein in the scattering. 

The neutrons scatter from the $H$ at points $r_{i}(t)$ in the protein. The incoherent DSF is proportional 
to an average over the N self correlation functions of the $r_{i}(t)$ of each individual H. The H is 
distributed in very different environments throughout the protein and these correlation functions may be 
quite different with different time scales\cite{Meinhold:08,Kneller:09,Yi:11}. Our first approximation is 
to represent this average over all H by a single representative H so that the $I_{inc}(\textbf{Q},t)$ 
reduces to
\begin{equation} \label{e7}
 I(\textbf{Q},t) = \langle \exp(-i\textbf{Q}\cdot\textbf{r}(t)) \exp(i\textbf{Q}\cdot\textbf{r}(0))\rangle .
\end{equation}
This self correlation function of $\textbf{r}(t)$ represents a weighted distribution over correlation 
functions and will contain all time scales, short and long, of H throughout the protein. We also drop the 
incoherent ``inc" from \sqw and $I(\textbf{Q},t)$ from now on. Eq.~(\ref{e7}) is the usual approximation 
made in the analysis of experimental data.  

The neutron instrument has an energy resolution function, denoted here by $R(\omega)$. A perfect energy 
resolution would be $R(\omega) = \delta (\omega)$, no resolution width. A representation of an $R(\omega)$ 
of finite width $W$ is

\begin{equation} \label{e8}
 R(\omega)= \frac{1}{\pi} \frac{W}{W^2+\omega ^2},
\end{equation}
a Lorentzian function, which has Fourier transform $R(t)=\exp (-Wt)$. We will also use a Gaussian \rqw. 
The observed function is a convolution of \sqw and \rw,

\begin{eqnarray} \label{e9}
 O(\textbf{Q},\omega) &=&  \int _{-\infty}^{\infty} d\omega'  S(\textbf{Q},\omega ') R(\omega -\omega ') 
\\
   &=&  \frac{1}{2\pi} \int _{-\infty}^{\infty} dt \exp(i\omega t) I(\textbf{Q},t) R(t) \\
O(\textbf{Q},\omega) &=& \frac{1}{2\pi} \int _{-\infty}^{\infty} dt \exp(i\omega t) I(\textbf{Q},t) \exp 
(-Wt).
\end{eqnarray}  

Commonly measured is the elastic (zero energy transfer, $\omega$ = 0) component of the incoherent 
$O(\textbf{Q},\omega)$,
\begin{equation} \label{e10}
 O(\textbf{Q},\omega = 0) =  \frac{1}{2\pi} \int _{-\infty}^{\infty} dt I(\textbf{Q},t) \exp (-Wt).
\end{equation}  

\begin{table}[htbp]
\caption{\footnotesize
The resolution width and the resolution time of neutron scattering instruments.}
\begin{tabular}{lllr}
\hline
\hline
Instrument & Energy Resolution   & $W$ & Time Scale  \\
 & ($\mu$eV) & ($$THz) & ($$ps) \\ 
\hline
\\
IN16  & $\sim$1  & 0.00025 & $\sim$4000  \\
\\
IN10  & $\sim$1  & 0.00025 & $\sim$4000  \\
\\
IN13  & $\sim$10  & 0.0025 & $\sim$400  \\
\\
IN6  & $\sim$100  & 0.025 & $\sim$40  \\
\\
IN5  & $\sim$100  & 0.025 & $\sim$40  \\
\\
HER  & $\sim$100  & 0.025 & $\sim$40  \\
\\
GP-TAS  & $\sim$1000  & 0.25 & $\sim$4  \\
\hline
\hline
\end{tabular}
\end{table} 

In $O(\textbf{Q},\omega = 0)$, we see that the role of a finite resolution width is to cut off the 
integrand after at time $\tau \sim W^{-1}$ so that long time processes in $I(\textbf{Q},t)$ are not 
observed in $O(\textbf{Q},\omega = 0)$. The higher the instrument resolution, the longer the time in 
$I(\textbf{Q},t)$ that can be observed in $O(\textbf{Q},\omega = 0)$. This is completely general 
independent of the form of $I(\textbf{Q},t)$ and $R(t)$. Only for infinitely sharp resolution, $W$ = 0, 
$R(t) = 1$ and $R(\omega) = \delta (\omega)$, are all long time motions in $I(\textbf{Q},t)$ observed in 
$O(\textbf{Q},\omega = 0)$.

In experiments, it is often assumed that the observed $O(\textbf{Q},\omega = 0)$ is given by 
$O_{exp}(\textbf{Q},\omega = 0) = \exp [-\frac{1}{3} Q^2 \langle r^2\rangle]$. The experimental $\langle 
r^2\rangle _{exp}$ is then obtained as $\langle r^2\rangle _{exp} = -3{d \ln O_{exp}(\textbf{Q},\omega 
=0)}/{dQ^2}$ as in Eq.~(\ref {e1}).

Since from Eq.~(\ref {e10}) the observed $O(\textbf{Q},\omega = 0)$ depends on the instrument resolution 
width $W$, this $\langle r^2\rangle _{exp}$ will depend on the instrument resolution.

\begin{figure*}
\includegraphics[scale=0.22,angle=0]{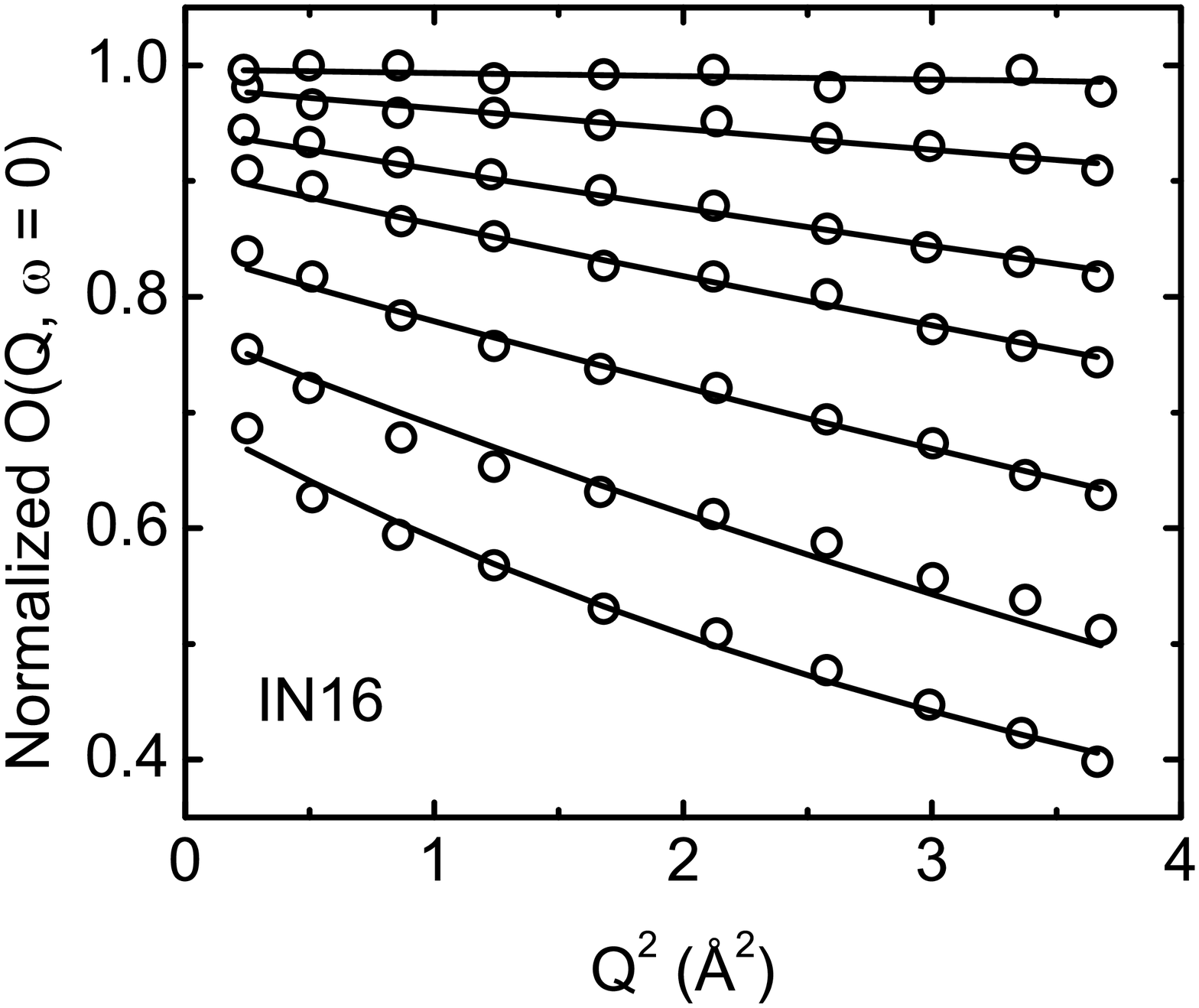}
\includegraphics[scale=0.22,angle=0]{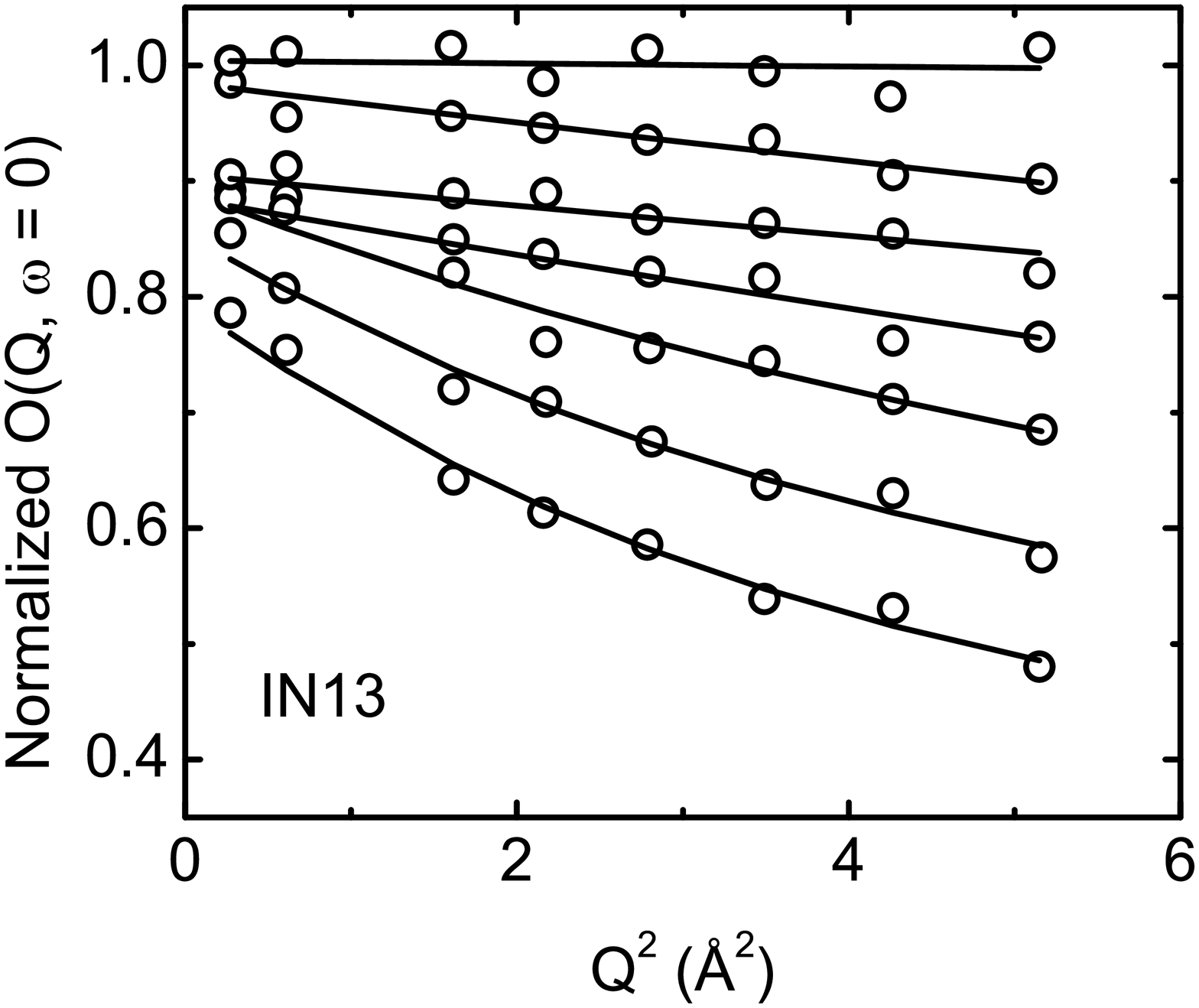}
\includegraphics[scale=0.22,angle=0]{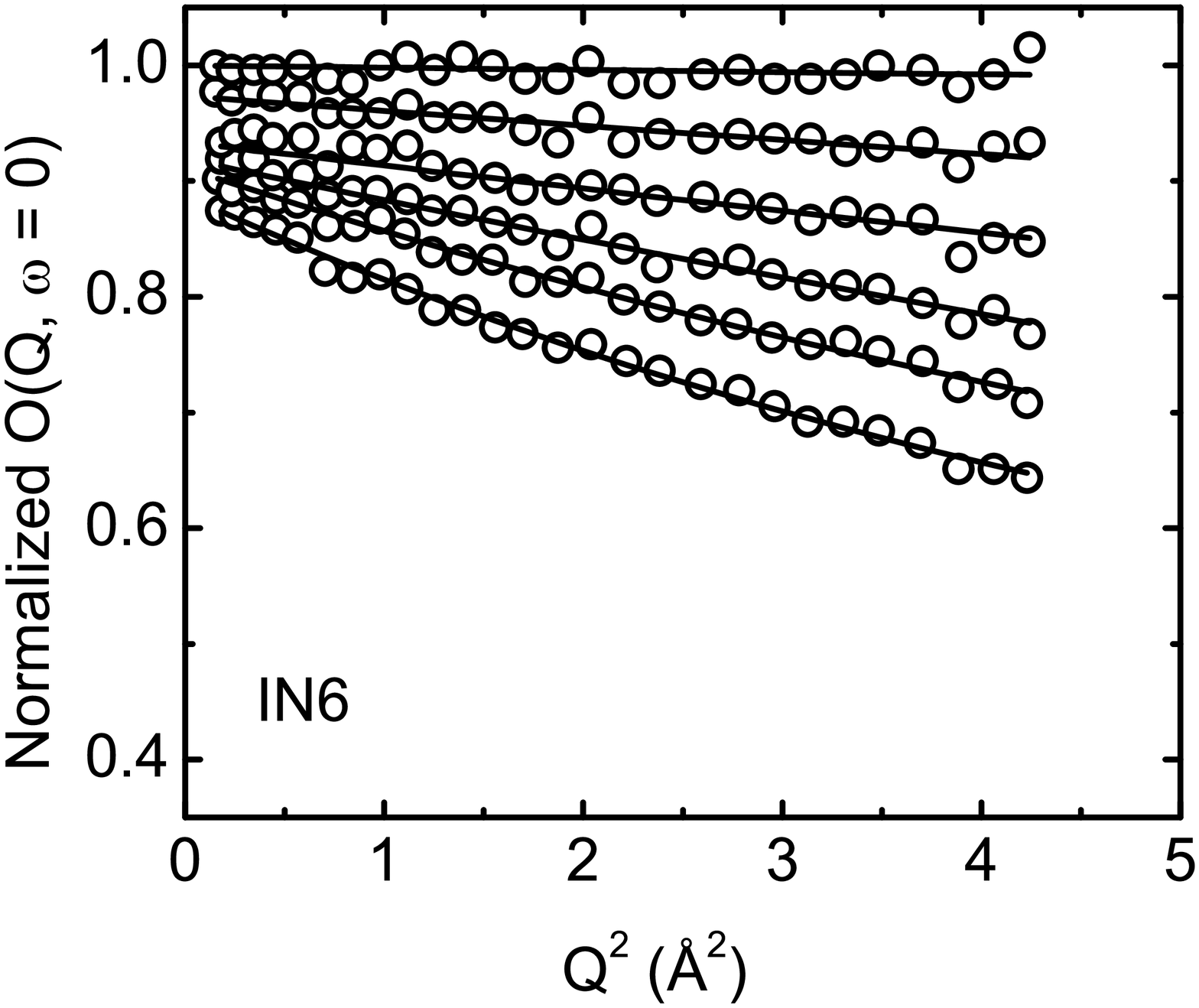}\\
\includegraphics[scale=0.22,angle=0]{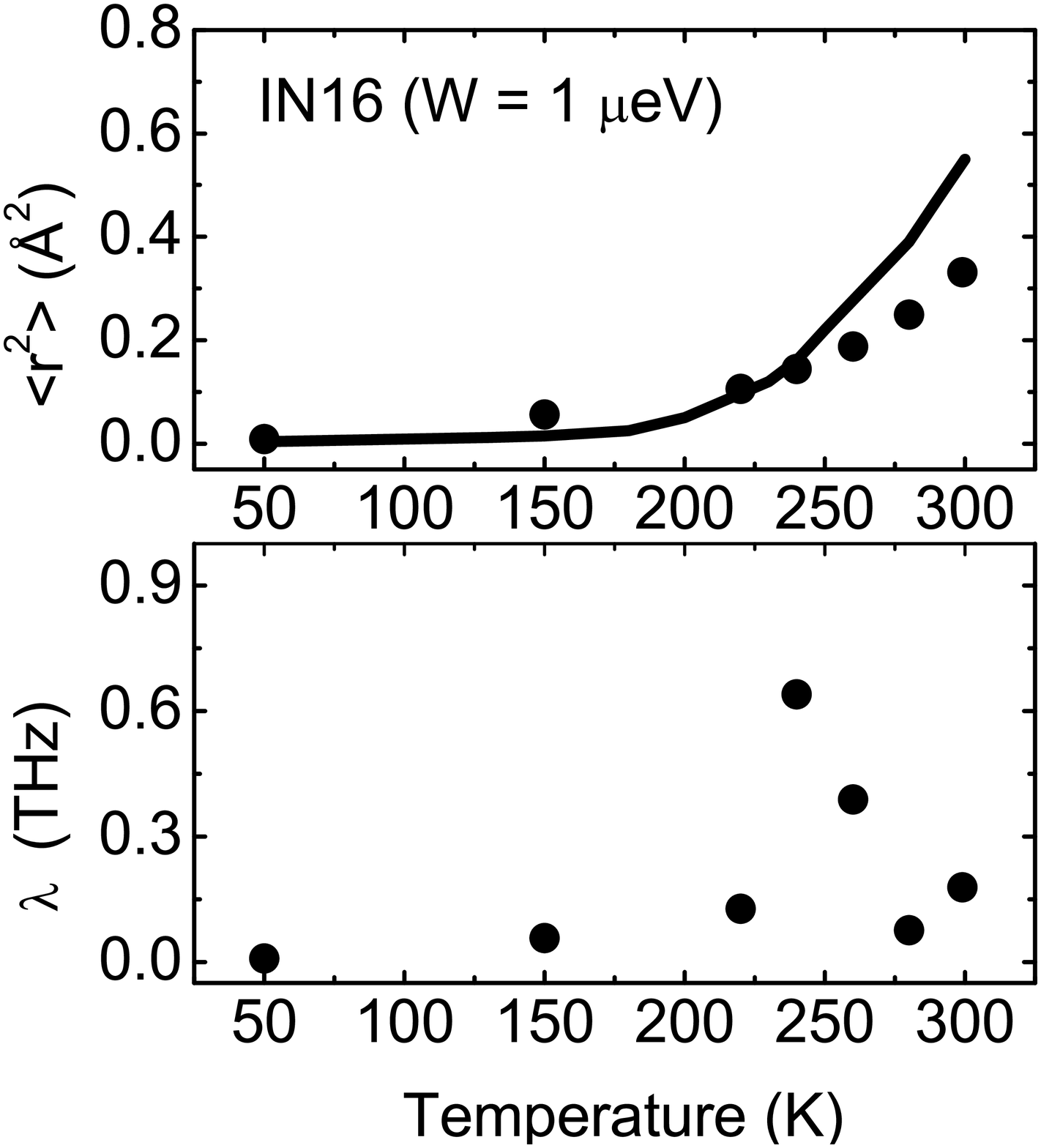}
\includegraphics[scale=0.22,angle=0]{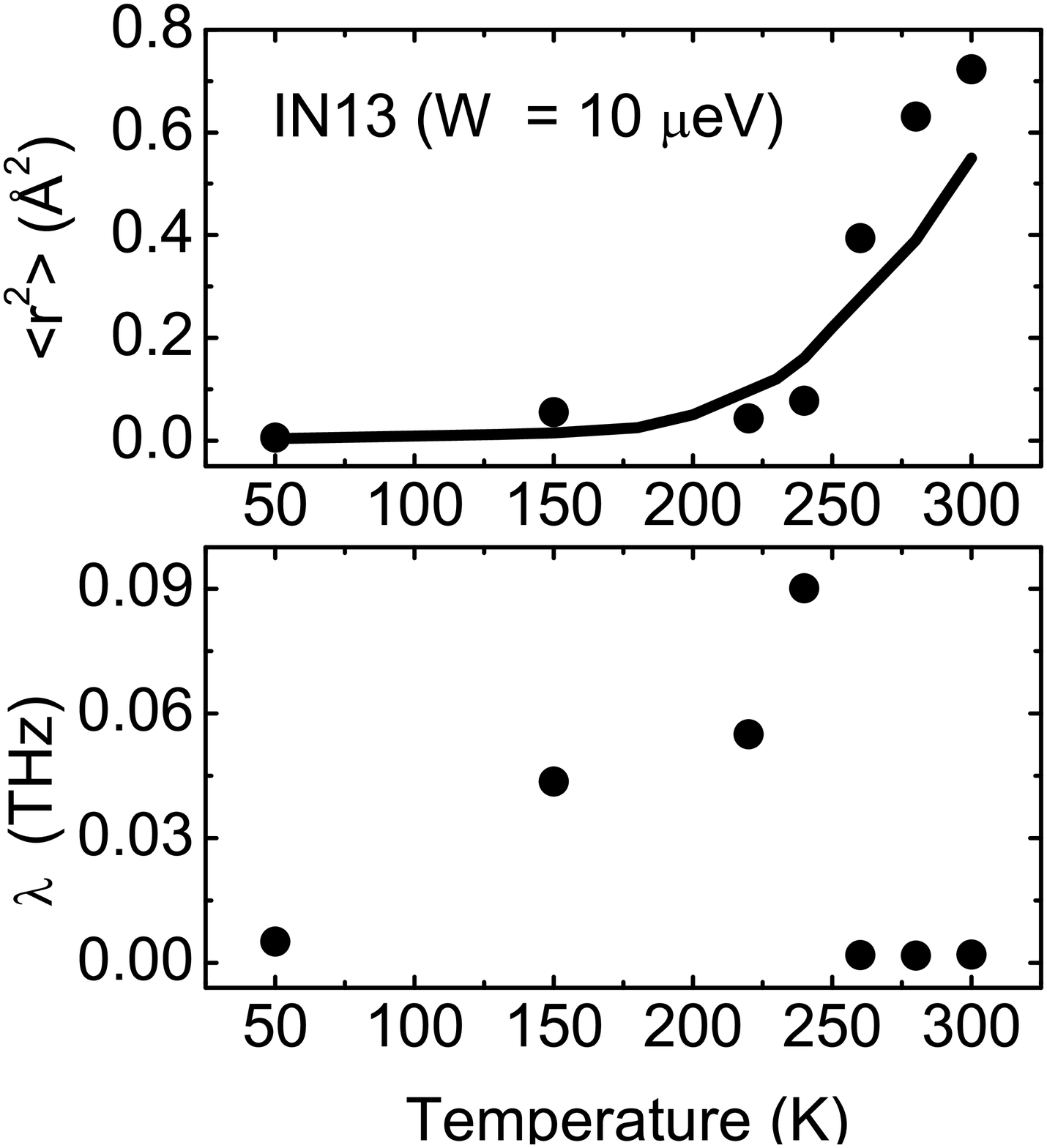}
\includegraphics[scale=0.22,angle=0]{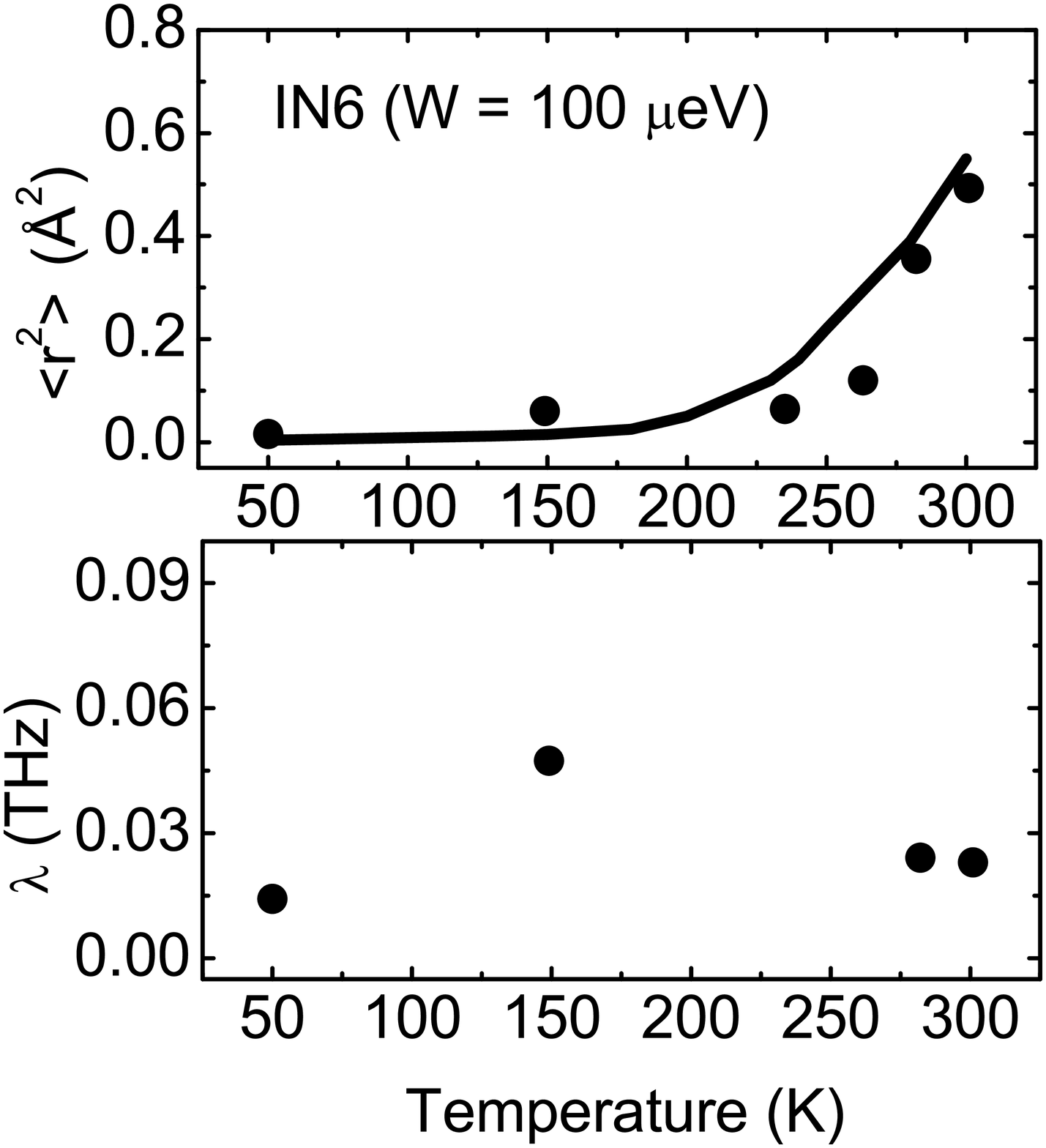}\\
\vspace{0.5cm}
\caption{Upper frame: Elastic DSF, $O_N(\textbf{Q}, \omega = 0)$, of $H$ in heparan sulphate observed by 
Jasnin et.al\cite{Jasnin:10} (open circles) and fit of model Eq.~(\ref{e2}) to the observed 
$O_N(\textbf{Q}, \omega = 0)$. Lower frames: the best fit values of the intrinsic $\langle r^2\rangle$ and 
$\lambda$. The solid line is a guide to the eye for $\langle r^2\rangle$, the same line for all three 
instruments.}
\label{f2}
\end{figure*}

\subsection{The Model}

The aim is to construct a simple model for $I(\textbf{Q},t)$ which we can fit to data to obtain the 
intrinsic value of $\langle r^2\rangle$ that includes the contributions from all motions, including long 
time motions.

We begin by separating $I(\textbf{Q},t)$ into a $t = \infty$ ($I_{\infty} = I(\textbf{Q},t = \infty)$) and 
a time dependent part,
\begin{equation} \label{e12}
I(\textbf{Q},t) = I_{\infty} + (I(\textbf{Q},t)-I_{\infty}).
\end{equation}
where, from Eq.~(\ref{e7}), 
\begin{eqnarray} \label{e13}
I_{\infty} = I(\textbf{Q},t = \infty) &=& \langle \exp(-i\textbf{Q}\cdot\textbf{r}(\infty)) 
\exp(i\textbf{Q}\cdot\textbf{r}(0))\rangle \nonumber \\
   &=& \langle \exp(-i\textbf{Q}\cdot\textbf{r}(\infty)) \rangle \langle 
\exp(i\textbf{Q}\cdot\textbf{r}(0))\rangle \nonumber \\
   &=& \langle \exp(-i\textbf{Q}\cdot\textbf{r}(0)) \rangle \langle \exp(i\textbf{Q}\cdot\textbf{r}(0))\rangle 
\nonumber \\
   &=& \exp(- \frac{1}{3}Q^2 \langle r^2 \rangle + O(Q^4))
\end{eqnarray}
is the infinite time limit. To obtain the last expression we assume: (1) that $\textbf{r}(\infty)$ and 
$\textbf{r}(0)$ are completely uncorrelated so that the averages of them are independent, (2) that the 
system is translationally invariant in time (no CM motion) so that  $\textbf{r}(\infty) = \textbf{r}(0)$ 
and (3) that in a cumulant expansion of $\langle \exp (-i\textbf{Q}\cdot\textbf{r})\rangle$, cumulants beyond 
the second cumulant are negligible. The latter is valid if $\textbf{Q}$ is small or if the distribution 
over $\textbf{r}$ is approximately a Gaussian distribution. The cumulants beyond the second vanish exactly 
for all $\textbf{Q}$ if the distribution over $\textbf{r}$ is exactly Gaussian. 

We take $\langle r^2\rangle$ in $I(\textbf{Q}, \infty)$ in Eq.~(\ref{e13}) as the definition of the 
intrinsic, long time value of $\langle r^2\rangle$ in the protein that includes all motional processes.  
It is the $\langle r^2\rangle$ that would be observed with an infinitely high resolution instrument, 
$R(\omega) = \delta(\omega)$. That is, $I_{\infty} = \exp (-Q^2\langle r^2\rangle/3)$ is the definition of 
the intrinsic $\langle r^2\rangle$ as in Eq.~(\ref{e3}).


The time dependent part of $I(\textbf{Q},t)$ has the limits
 \begin{displaymath}
   I'(\textbf{Q},t) = I(\textbf{Q},t)-I_{\infty}  = \left\{
     \begin{array}{lr}
       1-I_{\infty} & t=0 \\
       0   & t = \infty
     \end{array}
   \right.
\end{displaymath} 
We model this by the function
\begin{equation} \label{e15}
I'(\textbf{Q},t) = (1-I_{\infty}) C(t)
\end{equation}
where $C(t)$ has the limits $C(t=0)=1$, $C(t=\infty)=0$. An example is 
\begin{equation} \label{ct}
C(t) = \exp (-\lambda t),
\end{equation}
where $\lambda$ represents the decay constant of correlations in the protein. The model is

\begin{equation} \label{e16}
I(\textbf{Q},t) = I_{\infty}(\textbf{Q}) + (1- I_{\infty}(\textbf{Q}))C(t).
\end{equation}
which is constructed to have the correct limits at $t=0$ and $t=\infty$ and to have a plausible 
representation of a single motional decay process, for example diffusion, at intermediate times.

Much more accurate description of motional processes that incorporate a spectrum of relaxation times have
been developed and implemented.\cite{Kneller:05,Kneller:07,Calandrini:08} This leads to more complete 
expressions 
for $C(t)$, especially at long times. Specific examples are a stretched exponential and particularly the 
fractional
Ornstein-Uhlenbeck motional processes which provides relaxation functions the include long time motions. 
However, as discussed
further below, in this application of extracting the intrinsic $\langle r^2 \rangle$ from fits to 
experimental data we found 
that the intrinsic $\langle r^2 \rangle$ obtained was not sensitive to the form of $C(t)$ used in the 
model Eq. (17). 

Substituting the model $I(\textbf{Q},t)$ in $O(\textbf{Q},\omega = 0)$ the observed elastic function is 
 
\begin{eqnarray} \label{e17}
 O(\textbf{Q},\omega = 0) &=& \frac{1}{2\pi} \int _{-\infty}^{\infty} dt  I(\textbf{Q},t) \exp(-Wt) 
\nonumber \\
   &=&   I_{\infty} \frac{1}{\pi W}+ (1-I_{\infty})\frac{1}{\pi (W+\lambda )}.
\end{eqnarray}  
The data is usually presented as the ``normalized" $O(\textbf{Q},\omega = 0)$, the $O(\textbf{Q},\omega = 
0)$ above divided by $O(\textbf{Q} = 0,\omega = 0)$ at $\textbf{Q} = 0$. From Eq.~(\ref{e17}), the model 
$O(\textbf{Q} = 0,\omega = 0) = (\pi W)^{-1}$ and  $O_N(\textbf{Q},\omega = 0)$ is given by Eq. (2).
A constant $A$ is added to Eq.~(\ref{e2}) since the data for different $\textbf{Q}$ values are sometimes 
separated from one another in a figure by a constant for clarity. 

The $ O_N(\textbf{Q},\omega = 0) $ is the simple model that we fit to data for a given $W$. The $\langle 
r^2\rangle$, $\lambda$ and $A$ are treated as free fitting parameters to be determined by the best fit to 
data. The parameter $A$ plays no role. In this way we determine the intrinsic long time value of $\langle 
r^2\rangle$ and $\lambda$. The intrinsic \rs~ should be independent of $W$. It can be compared with the 
experimental values $\langle r^2\rangle _{exp}$ which are obtained from the slope of the observed 
$O_N(\textbf{Q},\omega = 0) $ using (12) and which are resolution dependent. Eq.~(\ref{e2}) is the initial 
simple model of $O_N(\textbf{Q},\omega = 0)$ we use in section 3 to test the method.

We can also obtain a value of the MSD using the slope expression, Eq.~(\ref{e1}), and our model $ 
O_N(\textbf{Q},\omega = 0) $ of Eq.~(\ref{e2}). This value, as with $\langle r^2\rangle _{exp}$, will 
depend on the instrument resolution $W$ since $ O_N(\textbf{Q},\omega = 0) $ depends on $W$. The model MSD 
obtained from the slope, $\langle r^2\rangle _{slope} = -3{d \ln O_N(\textbf{Q},\omega =0)}/{dQ^2}$ and 
differentiating Eq.~(\ref{e2}) is given by Eq.~(\ref{e4}).
We see immediately that $ \langle r^2\rangle _{slope} $ depends on the ratio of $W/\lambda$. If $W \ll 
\lambda$, then the instrument resolution is high enough to catch all the decay of $I(\textbf{Q},t)$ to 
$I_{\infty}$ and $ \langle r^2\rangle _{slope} \longrightarrow \langle r^2\rangle $, the intrinsic and 
full $ \langle r^2\rangle $. If the fit of $ O_N(\textbf{Q},\omega = 0) $ to data is good, so that 
$\lambda$ and $ \langle r^2\rangle $ in Eq.~(\ref{e4}) are well determined, then we expect    
the $ \langle r^2\rangle _{slope} $ to reproduce the observed $ \langle r^2\rangle _{exp} $ well.

\begin{figure}[htbp!]
\hspace{-0.1cm}
\vspace{-0.5cm}
\includegraphics[scale=0.29,angle=0]{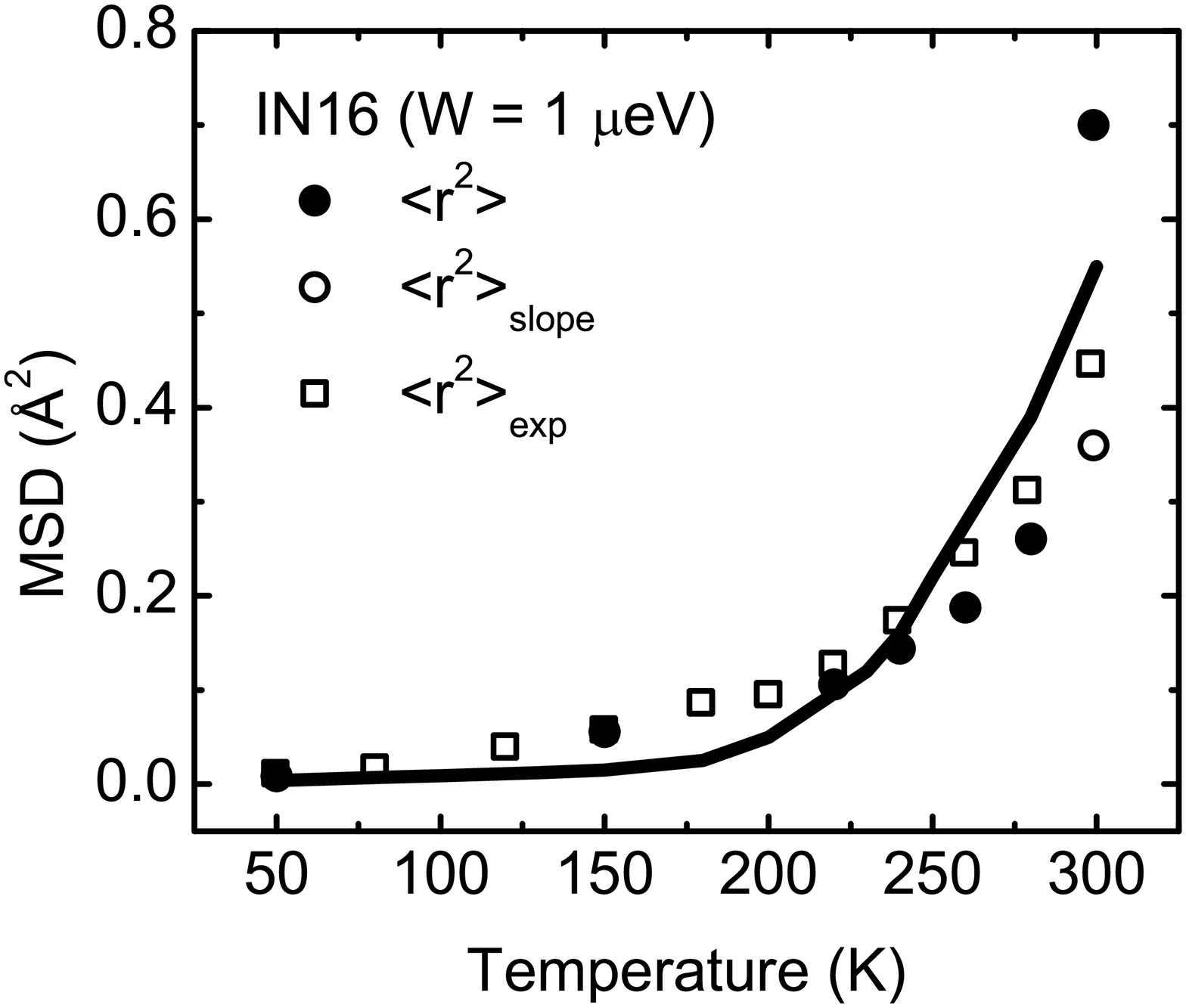}\\
\hspace{0.05cm}
\vspace{-0.5cm}
\includegraphics[scale=0.29,angle=0]{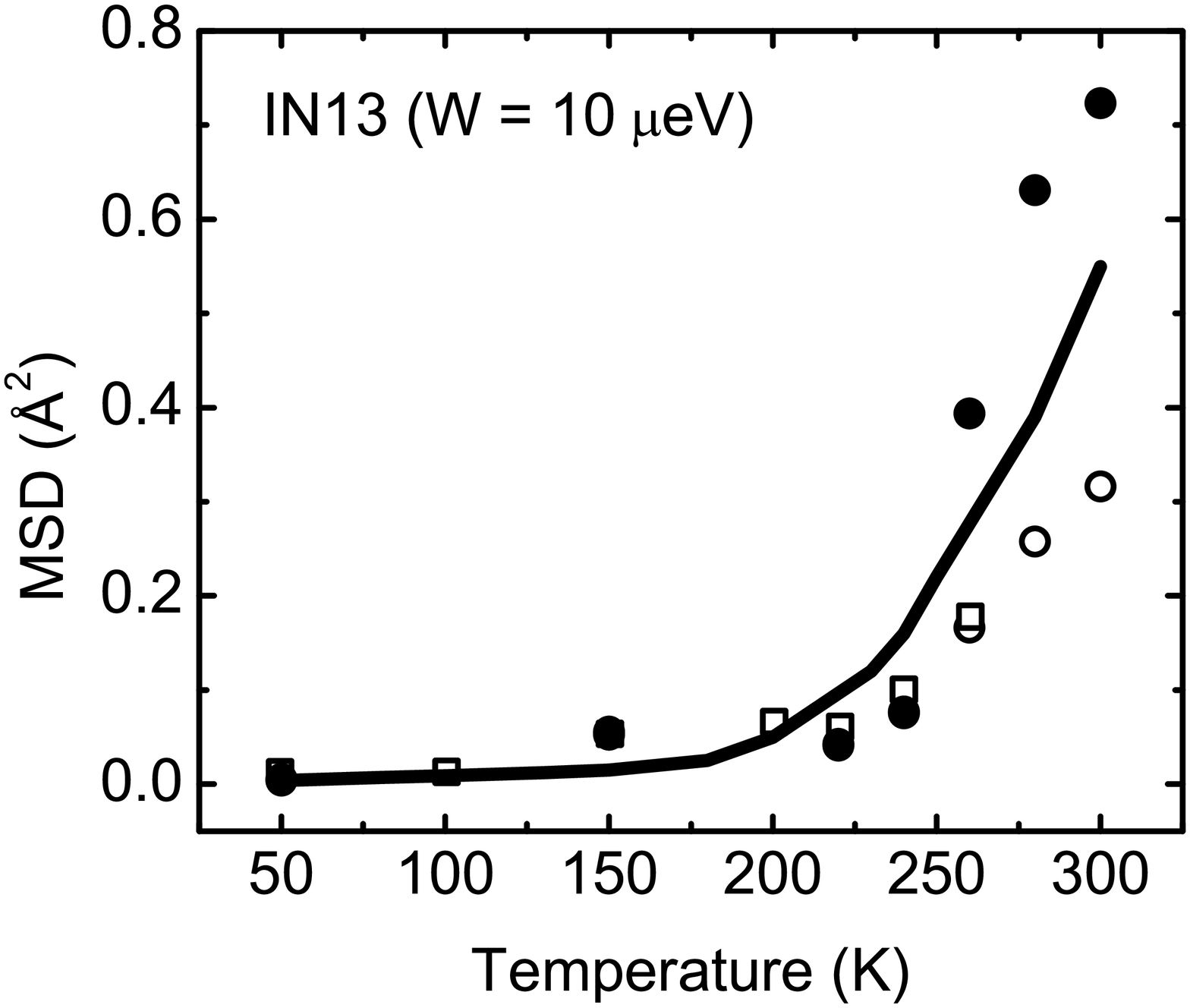}\\
\hspace{0.05cm}
\vspace{-0.5cm}
\includegraphics[scale=0.29,angle=0]{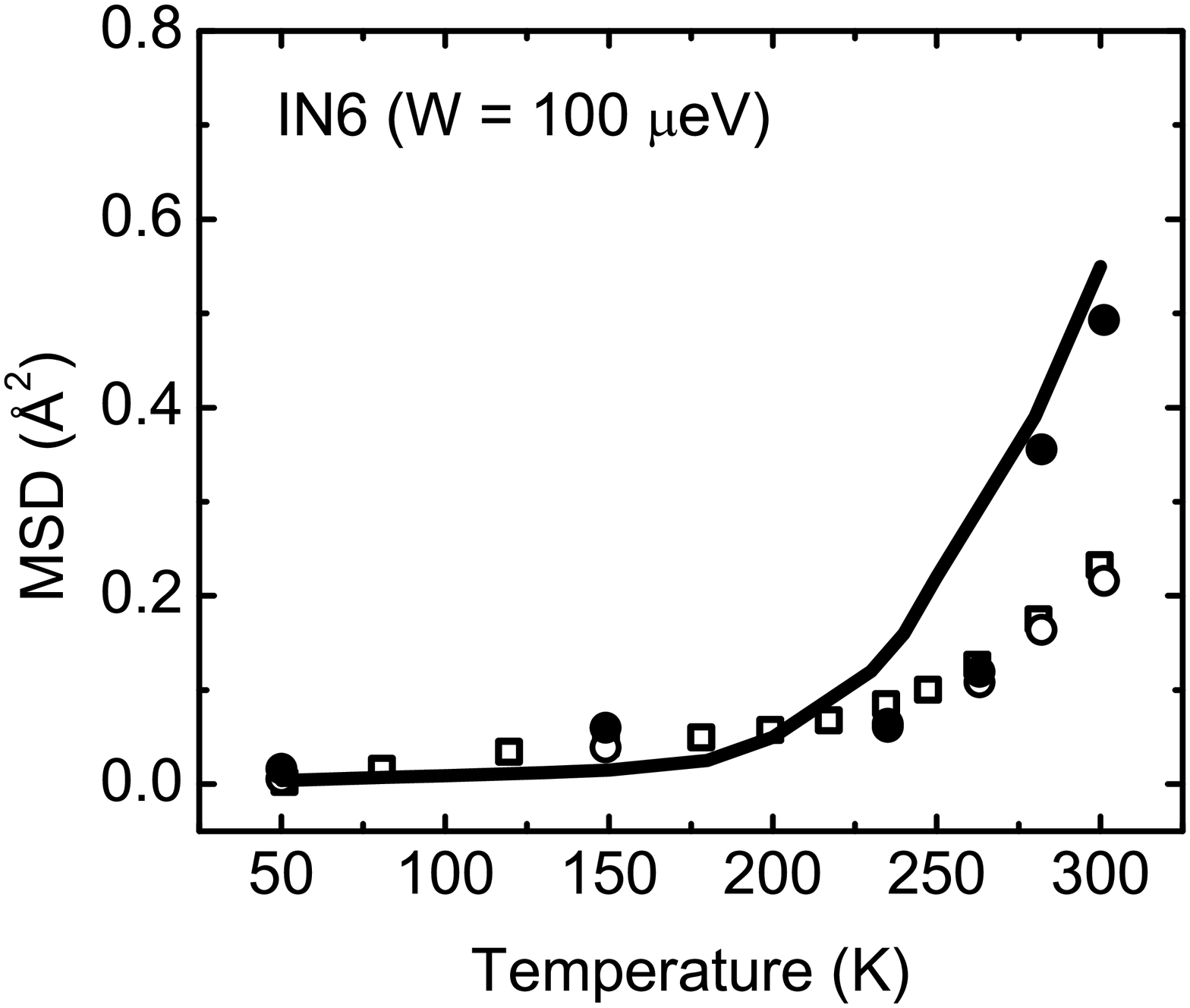}\\
\vspace{0.8cm}
\caption{MSD, \rsexp, in hydrated HS-0.4 observed Jasnin et. al\cite{Jasnin:10} on IN16, IN13 and IN6 
(open squares). The intrinsic $\langle r^2\rangle$ (solid circles) are the intrinsic MSD obtained by 
fitting the model Eq.~(\ref{e2}) to data shown in Fig.~\ref{f2}. The solid line is a guide to the eye 
through the intrinsic $\langle r^2\rangle$, the same for all instruments. The  $\langle r^2\rangle 
_{slope}$ (open circles) are the MSD calculated from Eq.~(\ref{e4}) which should be similar to \rsexp.}
\label{f3}
\end{figure}

\begin{figure*}
\hspace{-0.50cm}
\vspace{-0.5cm}
\includegraphics[scale=0.23,angle=0]{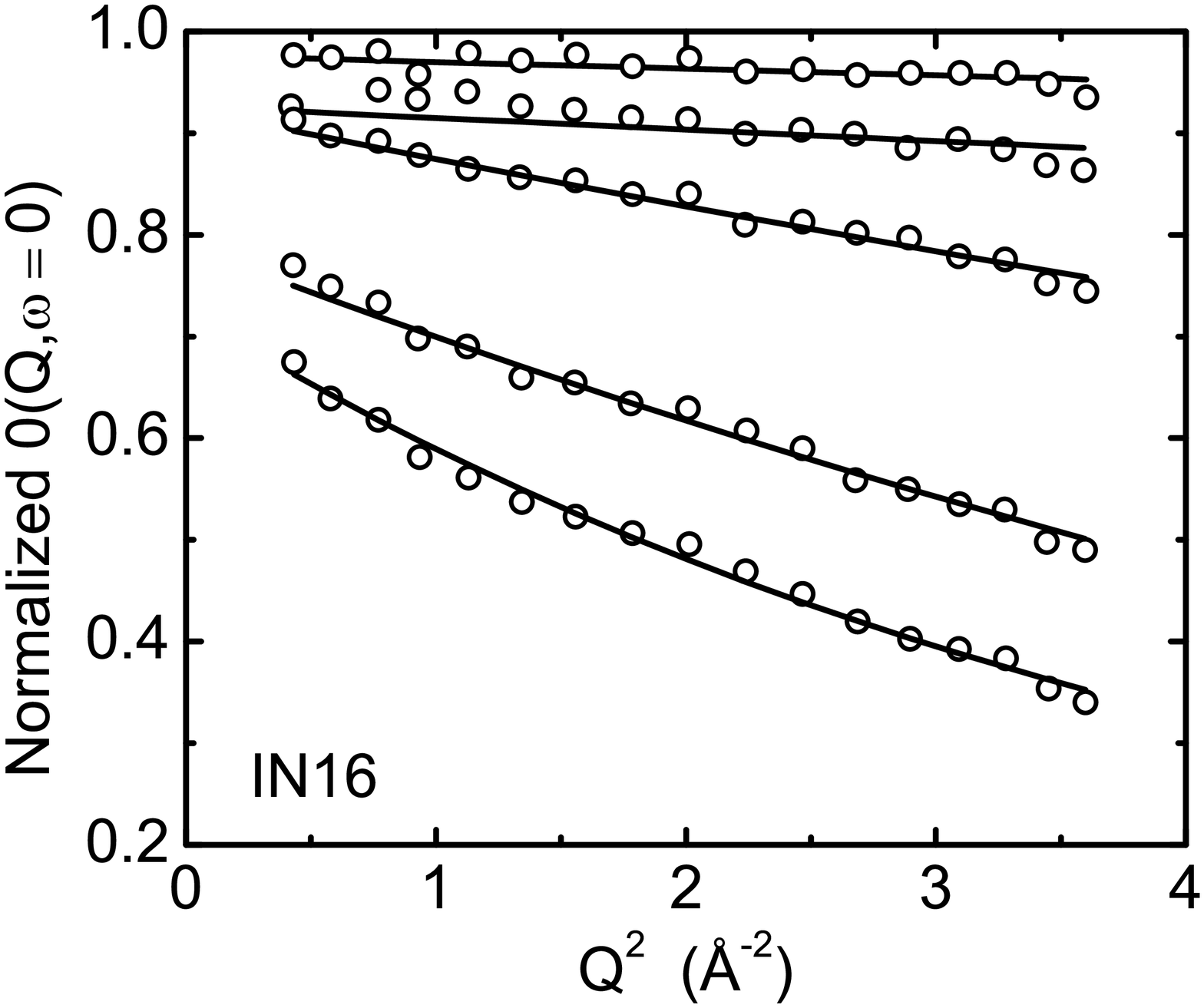}
\includegraphics[scale=0.23,angle=0]{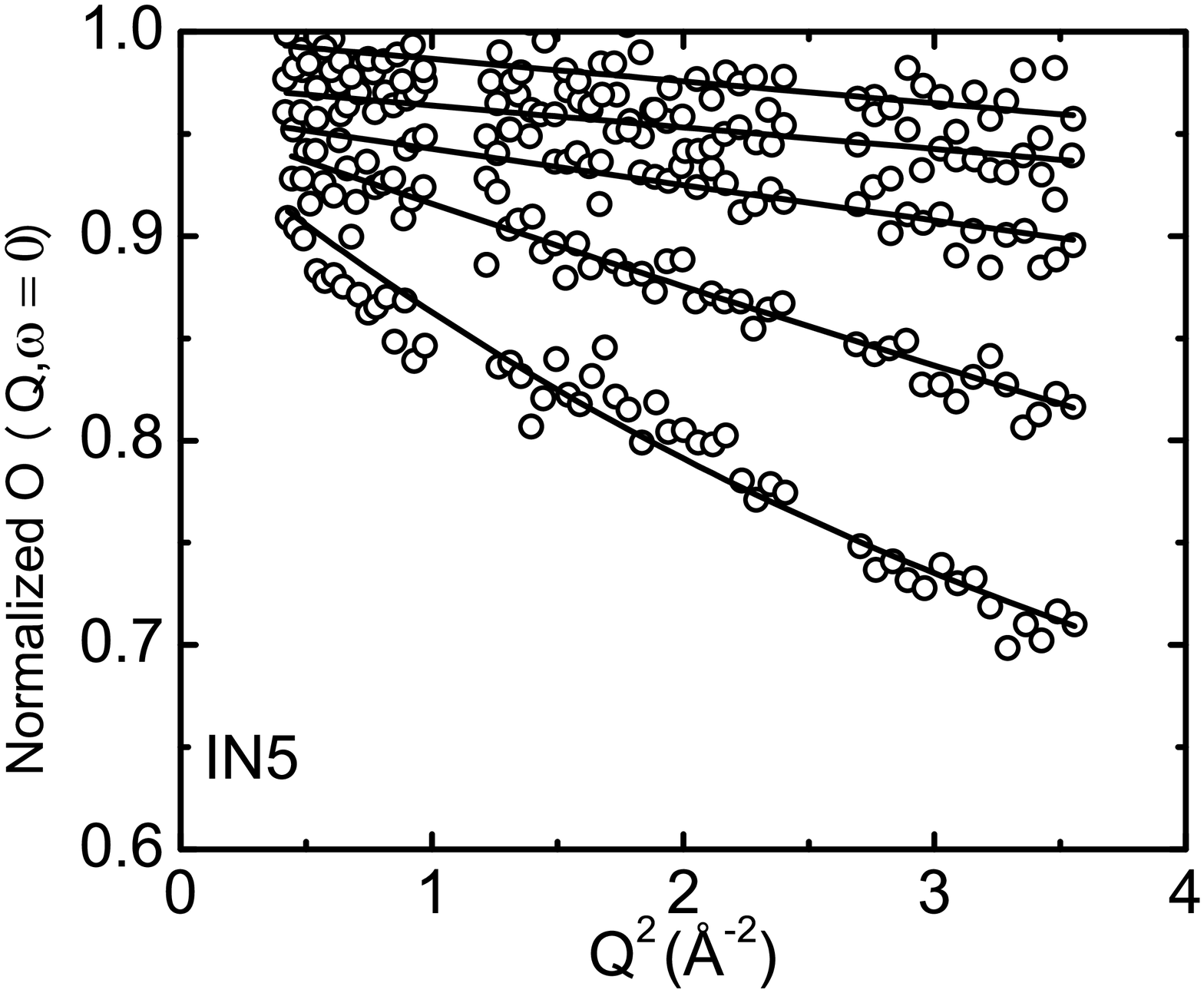}\\
\hspace{-0.50cm}
\vspace{-0.5cm}
\includegraphics[scale=0.23,angle=0]{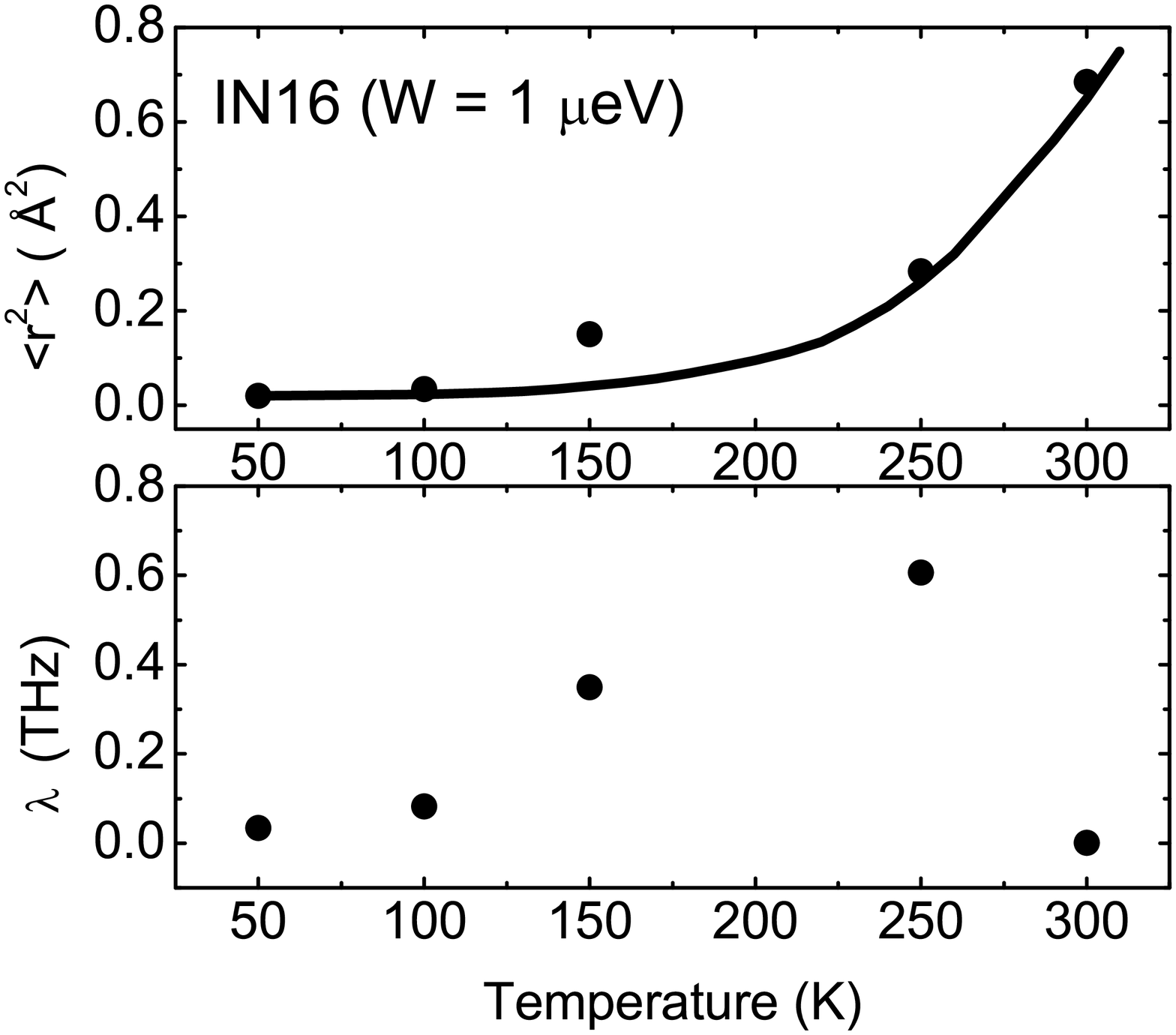}
\includegraphics[scale=0.23,angle=0]{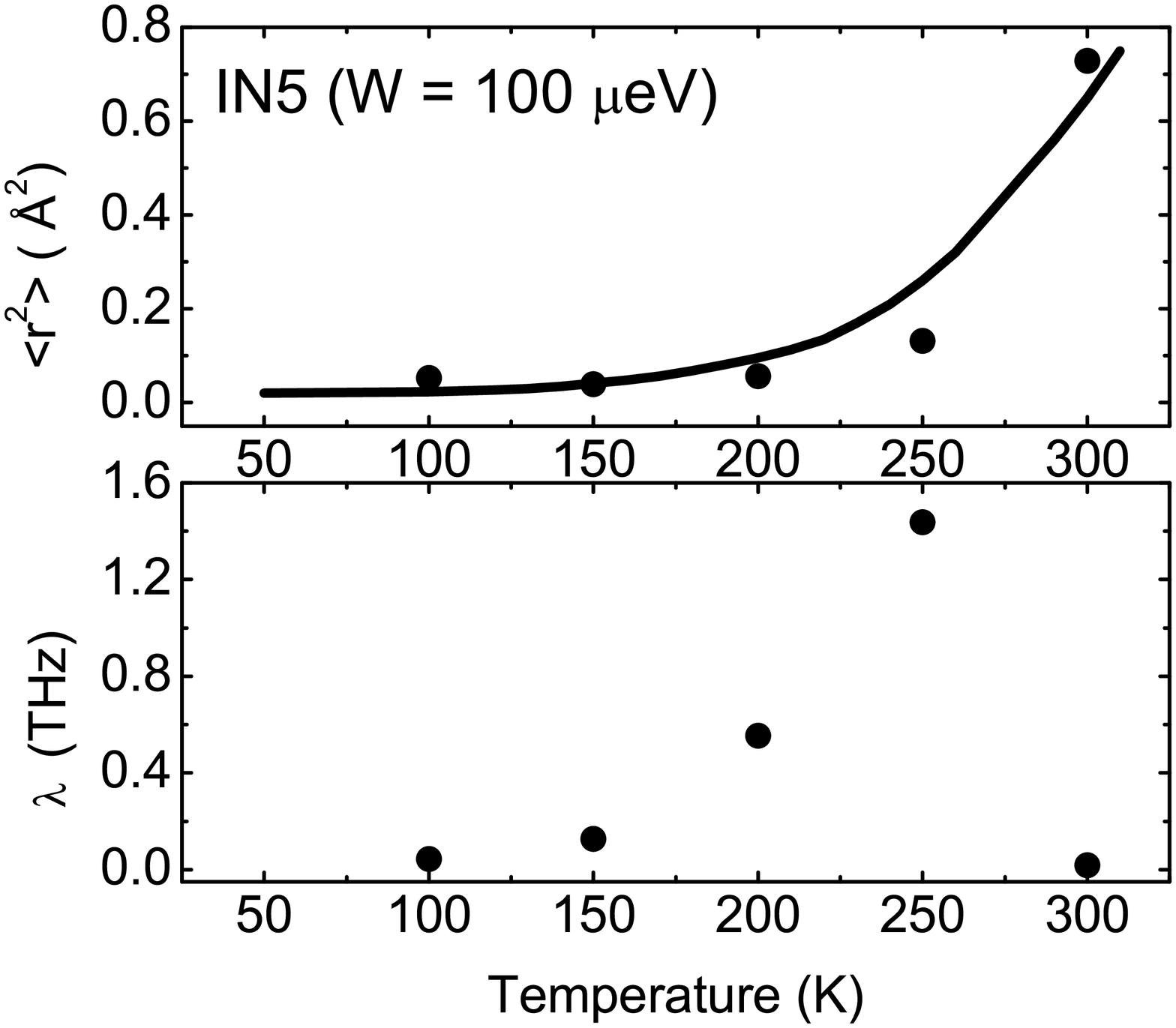}
\vspace{0.5cm}
\caption{Upper frame: Elastic DSF, $O_N(\textbf{Q}, \omega = 0)$, observed by Wood et. al\cite{Wood:08} on 
IN16 and IN5 (open circles) and fit of model Eq.~(\ref{e2}) to the observed $O_N(\textbf{Q}, \omega = 0)$. 
Lower frames: the best fit values of $\langle r^2\rangle$ and $\lambda$. The solid line is a guide to the 
eye to the best fit $\langle r^2\rangle$, the same line for both instruments.}
\label{f4}
\end{figure*}

\section{Results}

In this section we fit our model of the observed, resolution broadened elastic incoherent DSF 
$O_N(\textbf{Q},
\omega = 0)$ given by Eq.~(\ref{e2}) to data in the literature. The goal is to determine the intrinsic MSD 
$\langle r^2\rangle$ of $H$ in specific proteins and obtain a value for the relaxation parameter $\lambda$ 
which describes the approach of $\langle r^2\rangle$ to its long time, intrinsic value. 
%

\begin{figure}[t]
\hspace{-0.1cm}
\vspace{-0.5cm}
\includegraphics[scale=0.29,angle=0]{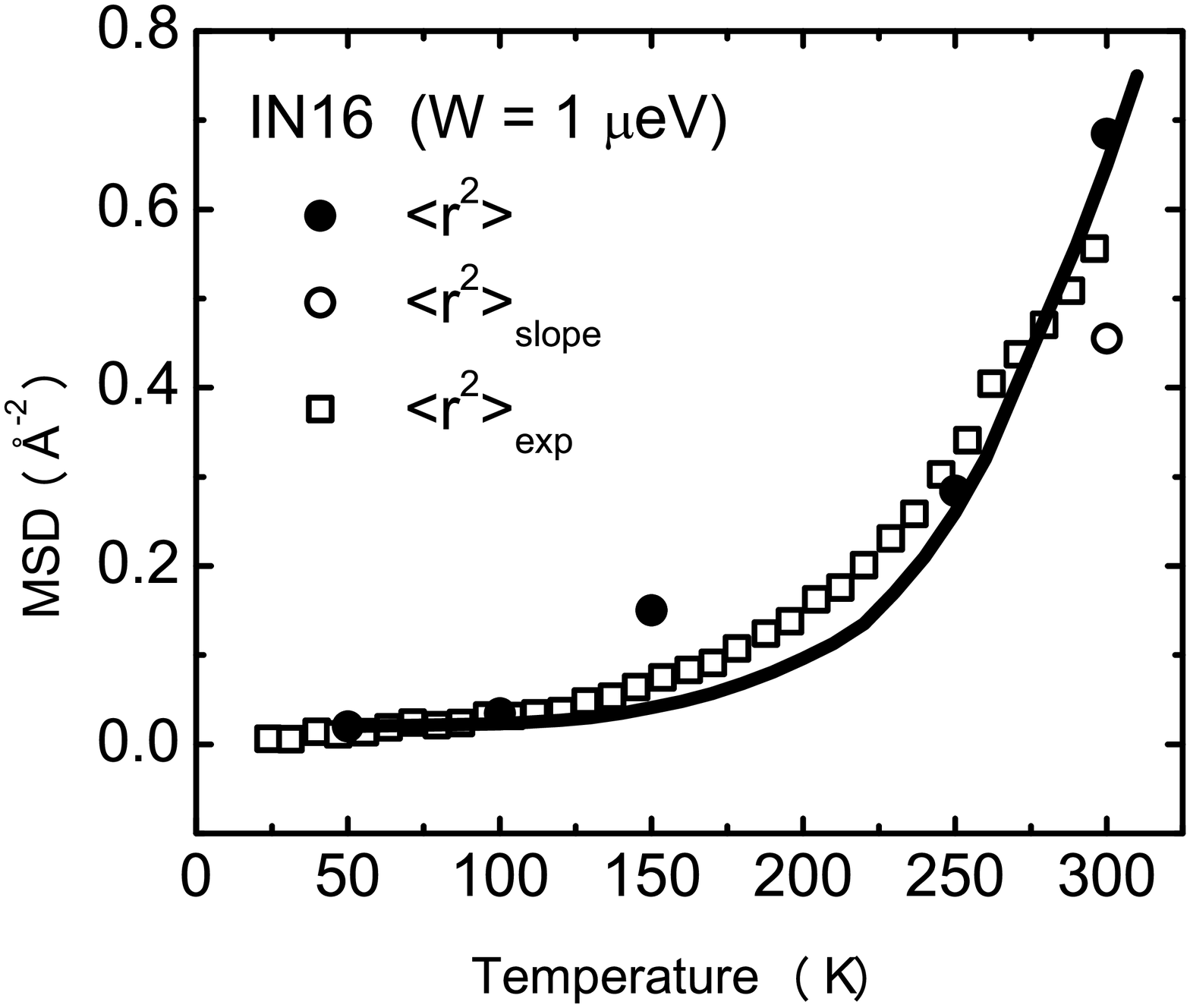}\\
\hspace{0.03cm}
\vspace{-0.5cm}
\includegraphics[scale=0.29,angle=0]{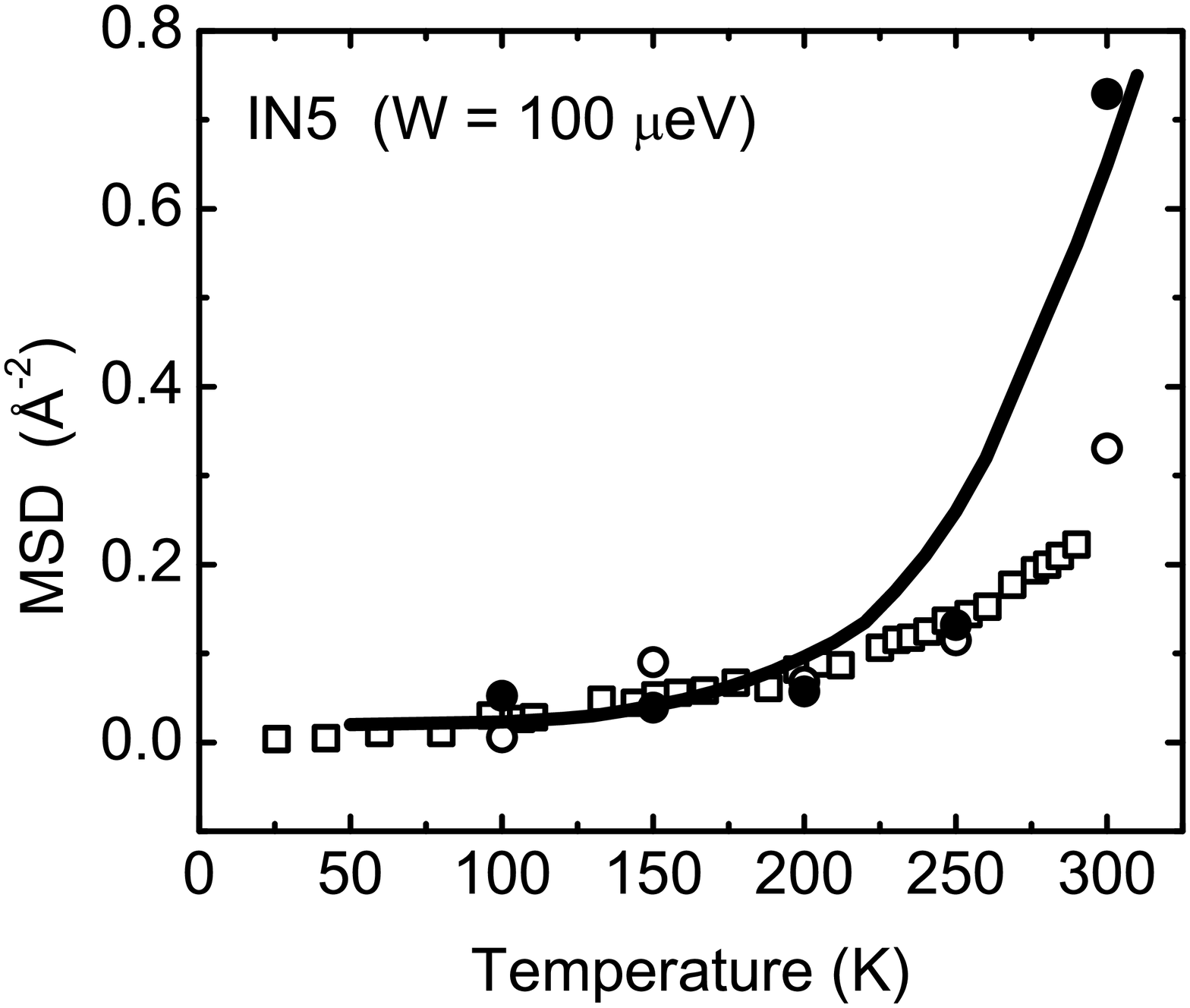}
\vspace{0.8cm}
\caption{MSD, \rsexp, in hydrated Ribonuclease A observed by Wood et. al \cite{Wood:08} on IN16 and IN5 
(open squares). The intrinsic MSD $\langle r^2\rangle$ (solid circles) are obtained by fitting the model 
Eq.~(\ref{e2}) to the data shown in Fig.~\ref{f4}. The solid line is a guide to the eye to the intrinsic 
MSD, the same for both instruments. The MSD $\langle r^2\rangle _{slope}$ (open circles) is calculated 
from Eq.~(\ref{e4}).}
\label{f5}
\end{figure}

\subsection{The MSD}

The top frame of Fig. 2 shows a fit of the model $O_N(\textbf{Q},\omega = 0)$ to the normalized  
$O(\textbf{Q},\omega = 0)$ observed in heparan sulphate (HS-0.4) by Jasnin et. al \cite{Jasnin:10}.  
$O(\textbf{Q},\omega = 0)$ is observed on three instruments which have different energy resolutions, $W = 
1$, $10$ and $100$ $\mu$eV. There is data for $O(\textbf{Q},\omega = 0)$ at seven temperatures, the lowest 
temperature at the top. The fits are generally good but some data points lie off the fitted line. The 
middle and bottom frames in Fig. 2 show the values of $\langle r^2 \rangle$ and $\lambda$ in the model 
$O_N(\textbf{Q},\omega = 0)$ that give the best fit. The best fit $\langle r^2 \rangle$ increases with 
temperature, markedly for temperatures above $T \simeq 230$ K. There is some scatter in the $\langle r^2 
\rangle$ which arises from the uncertainty in the fit. The scatter or uncertainty of $\lambda$ is 
particularly large showing that the data is relatively insensitive to the value of $\lambda$. Conversely, 
we can say that the data is not very discriminating or sensitive to the model employed for the time 
dependence of $C(t)$. However, importantly, the $\langle r^2 \rangle$ emerging from the fit is independent 
of the resolution, $W$. The solid black line in the middle frame of Fig. 2 is a guide to the eye through 
all the $\langle r^2 \rangle$, the same line for all instruments. This shows that while there is 
substantial fluctuations in the individual values of $\langle r^2 \rangle$, the average $\langle r^2 
\rangle$ obtained is independent of the instrument resolution $W$.

Similarly, Fig.~\ref{f3} compares the intrinsic \rs, the model \rsslope~ and the observed \rsexp~ in 
HS-0.4. The solid points are the intrinsic \rs~ emerging from the fits in Fig. \ref{f2} and the solid line 
is again a guide to the eye through these $\langle r^2 \rangle$, the same line for all three instruments. 
The $\langle r^2 \rangle _{exp}$ is the MSD obtained by Jasnin et. al from the slope of their data  using 
Eq.~(\ref{e1}). The $\langle r^2 \rangle _{slope}$ is obtained from the slope of the model 
$O(\textbf{Q},\omega = 0)$ given by Eq. (\ref{e4}). If the fit to the data is precise, the $\langle r^2 
\rangle _{slope}$ and $\langle r^2 \rangle _{exp}$ should agree. This is the case for IN13 and IN6 but 
less so for the IN16. When the instrument resolution $W$ is small, we expect $\langle r^2 \rangle 
_{slope}$ to coincide with the intrinsic $\langle r^2 \rangle$. This is the case for IN16 where $\langle 
r^2 \rangle _{slope}$ (open circles) lie on top of the $\langle r^2 \rangle$ (black dots). For IN16, the 
$\langle r^2 \rangle _{slope}$ and $\langle r^2 \rangle _{exp}$ differ somewhat for $T >$ 250 K. This 
indicates that there is not a good fit at higher temperatures, as can be seen in the upper frame of 
Fig.~\ref{f2}. The essential point of Fig.~\ref{f3} is that the intrinsic $\langle r^2 \rangle$ is 
independent of $W$ and that $\langle r^2 \rangle$ shows a marked increase at temperature, $T_D \simeq 230$ 
K, the intrinsic dynamical transition temperature of HS-0.4.
%
\begin{figure*}
\hspace{1.05cm}
\vspace{-0.55cm}
\includegraphics[scale=0.44,angle=0]{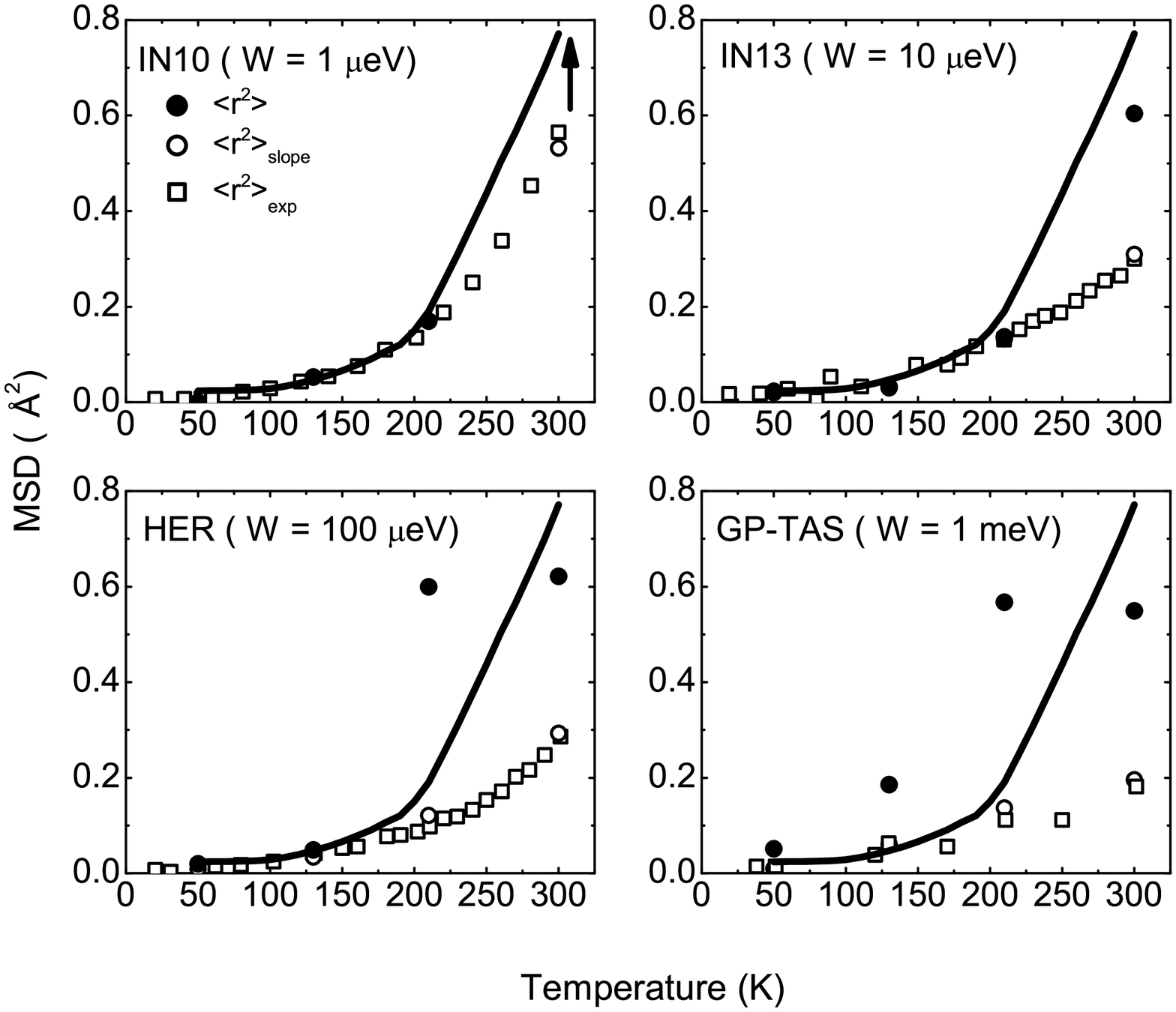}
\caption{MSD, \rsexp, of the hydrated SNase observed by Nakagawa et. al \cite{Nakagawa:10} on IN10 ($W = 
1~\mu$eV), IN13 ($W = 10~\mu$eV), HER ($W = 100~\mu$eV) and GP-TAS ($W = 1000~\mu$eV)(open squares). The 
intrinsic MSD $\langle r^2\rangle$(solid circle) is obtained by fitting the Eq.~(\ref{e2}) to the elastic 
intensity data observed Nakagawa et. al \cite{Nakagawa:10}. The solid line is a guide to the eye to the 
intrinsic MSD, the same for all instruments. The MSD $\langle r^2\rangle _{slope}$(open circles) is 
calculated from Eq.~(\ref{e4}).}
\label{f6}
\end{figure*}
%
\begin{figure*}
\hspace{2.01cm}
\vspace{-0.55cm}
\includegraphics[scale=0.44,angle=0]{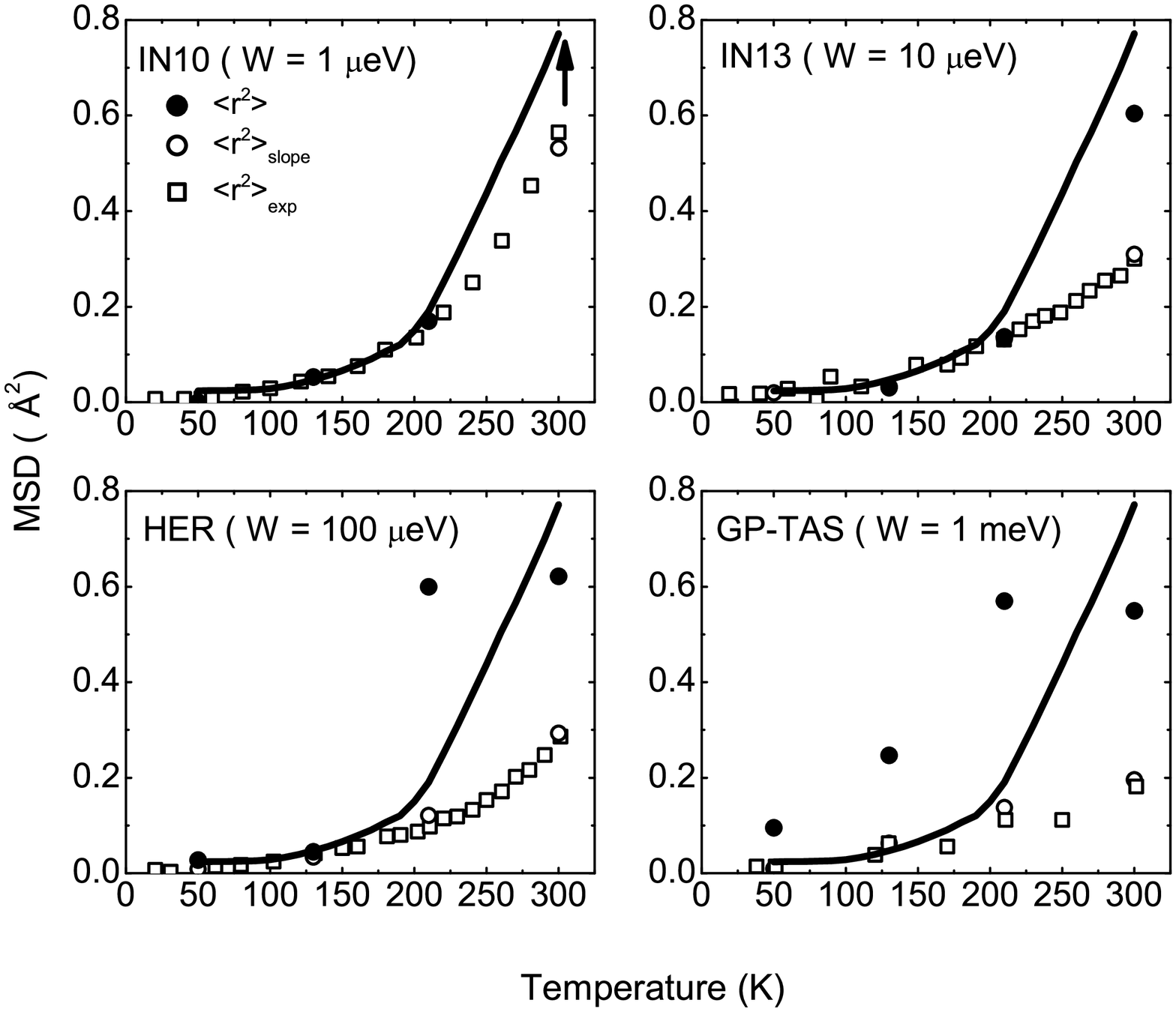}
\caption{MSD, \rsexp, of the hydrated SNase observed by Nakagawa et. al \cite{Nakagawa:10} on IN10 ($W = 
1~\mu$eV), IN13 ($W = 10~\mu$eV), HER ($W = 100~\mu$eV) and GP-TAS ($W = 1000~\mu$eV)(open squares). The 
intrinsic MSD $\langle r^2\rangle$(solid circle) is obtained by fitting the observed DSF equation, where 
the Gaussian resolution and $C(t)$ functions are used, to the elastic intensity data observed Nakagawa et. 
al \cite{Nakagawa:10}. The solid line is a guide to the eye to the intrinsic MSD, the same for all 
instruments. The MSD $\langle r^2\rangle _{slope}$(open circles) is calculated from the $Q^2$ dependence 
of the observed DSF.}
\label{f7}
\end{figure*} 
Similarly, the upper frame of Fig.~\ref{f4} shows the fit of the model $O_N(\textbf{Q},\omega = 0)$ of 
Eq.~(\ref{e2}) to the measured  $O_N(\textbf{Q},\omega = 0)$ in Ribonuclease A observed by Wood et. 
al\cite{Wood:08}. The fits are good except that there is significant scatter in the data taken on IN5. The 
best fit values of $\langle r^2 \rangle$ and $\lambda$ are shown in the lower frames of Fig.~\ref{f4}. 
Again the best fit values of the intrinsic $\langle r^2 \rangle$ is independent of the instrument energy 
resolution $W$, as seen from the solid line, a guide to the eye to the intrinsic $\langle r^2 \rangle$, 
the same for both instruments. In Fig.~\ref{f5}, the intrinsic $\langle r^2 \rangle$, the solid line, is 
compared with the resolution dependent observed $\langle r^2 \rangle _{exp}$ obtained from the slope of 
the data with $Q^2$. While the $\langle r^2 \rangle _{exp}$ is very different for the two instruments 
displaying the dependence of $\langle r^2 \rangle _{exp}$ on the resolution width $W$, the $\langle r^2 
\rangle _{exp}$ observed on IN16 agrees well with the intrinsic $\langle r^2 \rangle$. This indicates that 
$\langle r^2 \rangle$ in Ribonuclease A reaches its final, equilibrium value within a time period 
observable on IN16 ($\tau \sim 4$ nanoseconds). 
In a similar way, we fitted the model Eq.~(\ref{e2}) to values of $O_N(\textbf{Q},\omega = 0)$ observed by 
Nakagawa et. al\cite{Nakagawa:10} in Staphysloccal Nuclase (SNase) on four different instruments. The 
resulting intrinsic $\langle r^2 \rangle$ obtained from the fits are shown in Fig.~\ref{f6} as solid dots 
for each instrument with a solid line (guide) through the dots, the same line for all four instruments. 
Although there is substantial scatter in the $\langle r^2 \rangle$, an intrinsic $\langle r^2 \rangle$ 
independent of the instrument resolution width $W$ can be obtained by fitting the model to the observed 
$O_N(Q,\omega = 0)$.
To test the sensitivity of the results to the model of $C(t)$ and to the shape of the instrument 
resolution (IR), we changed $C(t)$ to Gaussian, $C(t) = \exp (-\lambda ^2 t^2/2)$, and the IR to Gaussian, 
$R(\omega) = \exp(-\omega ^2/2W^2)/\sqrt{2\pi W^2}$. With both $C(t)$ and $R(\omega)$ a Gaussian, an 
analytic expression for $O_N(\textbf{Q},\omega = 0)$ can again be obtained. Fig.~\ref{f7} shows the 
resulting intrinsic $\langle r^2 \rangle$ obtained from fitting the Gaussian model to the data of Nakagawa 
et. al. The $\langle r^2 \rangle$ in Figs.~(\ref{f6}) and (\ref{f7}) are barely distinguishable. This 
shows that intrinsic $\langle r^2 \rangle$ obtained is not sensitive to the form of $C(t)$ and $R(\omega)$ 
used in the model at the present level of precision of the data. 

\subsection{The Dynamical Transition Temperature, \td}

In this subsection, we consider the apparent dependence of the dynamical transition temperature on the 
instrument energy resolution $W$. The upper frame of Fig.~\ref{f8} shows the $\langle r^2 \rangle _{exp}$ 
of H in glutomate dehydrogenase observed on IN16 and IN6. The $\langle r^2 \rangle _{exp}$ is obtained as 
usual from the slope of the observed $O_{exp}(\textbf{Q},\omega =0)$ with $Q^2$ as given by 
Eq.~(\ref{e1}). In this example, the dynamical transition temperature, $T_D$, the temperature at which 
$\langle r^2 \rangle _{exp}$ increases markedly with temperature, appears to depend on the neutron 
instrument used. On IN16, the apparent $T_D$ is $T_D \simeq 150$K while that on IN6 is $T_D \simeq 230$K. 
To determine the intrinsic $\langle r^2 \rangle$ and $T_D$ for this protein, we fitted the model  $\langle 
r^2 \rangle _{slope}$ given by Eq.~(\ref{e4}) to the observed $\langle r^2 \rangle _{exp}$ in 
Fig.~\ref{f8}. Eq.~(\ref{e4}) is the model equivalent of Eq.~(\ref{e1}). Specially, at each temperature we 
determined the two parameters $\langle r^2 \rangle $ and $\lambda$ by setting $\langle r^2 \rangle 
_{slope}$ in Eq.~(\ref{e4}) equal to the $\langle r^2 \rangle _{exp}$ of Fig.~\ref{f8} for each 
instrument. The values of $\langle r^2 \rangle _{slope}$ actually fitted are shown as the solid lines 
through the data points in the upper frame. The resulting values of $\langle r^2 \rangle $ and $\lambda$ 
are shown as the lower frames of Fig.~\ref{f8}. 

\begin{figure}[ht!]
\hspace{0.03cm}
\vspace{-0.5cm}
\includegraphics[scale=0.29,angle=0]{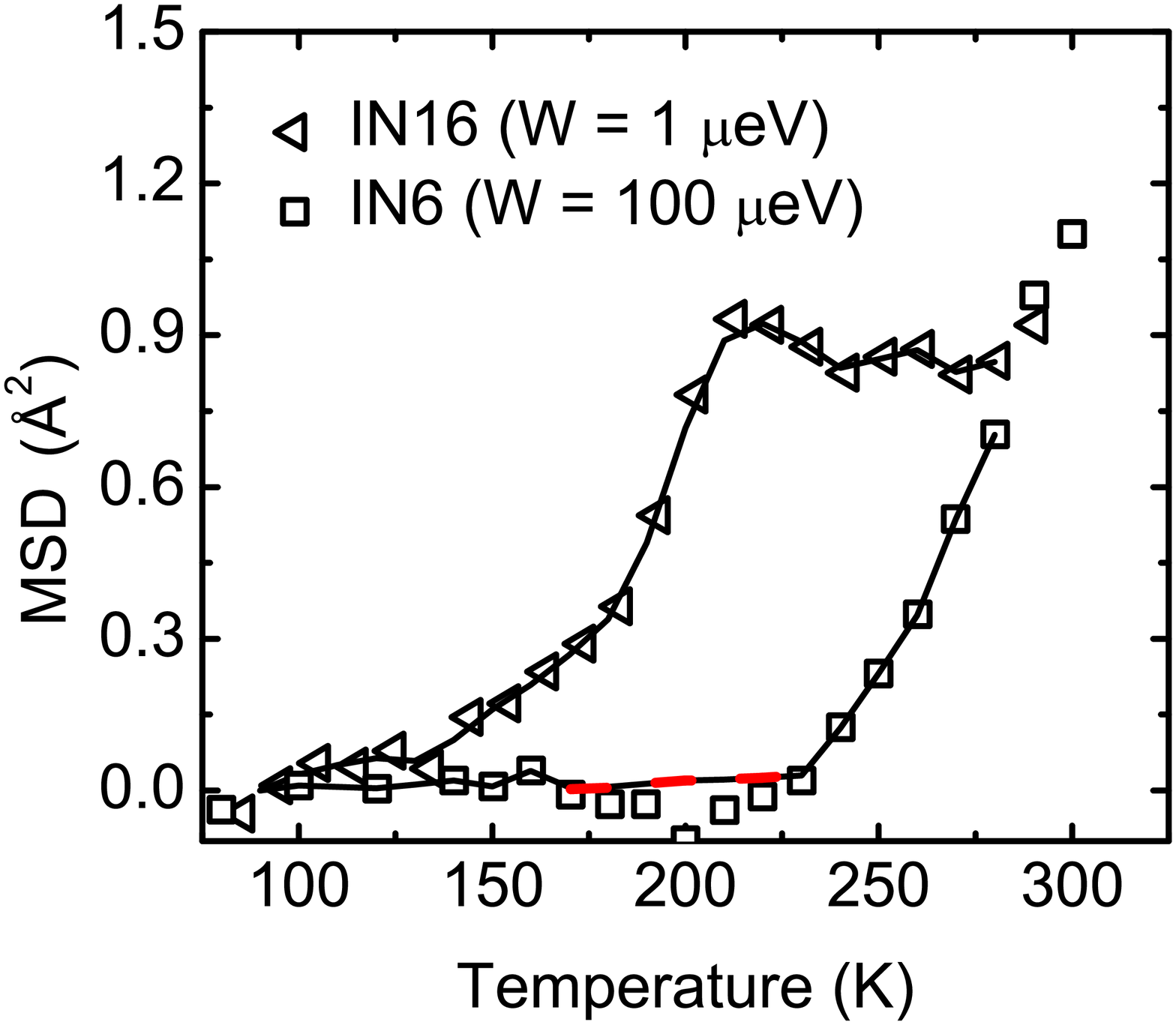}\\
\hspace{0.03cm}
\vspace{-0.5cm}
\includegraphics[scale=0.29,angle=0]{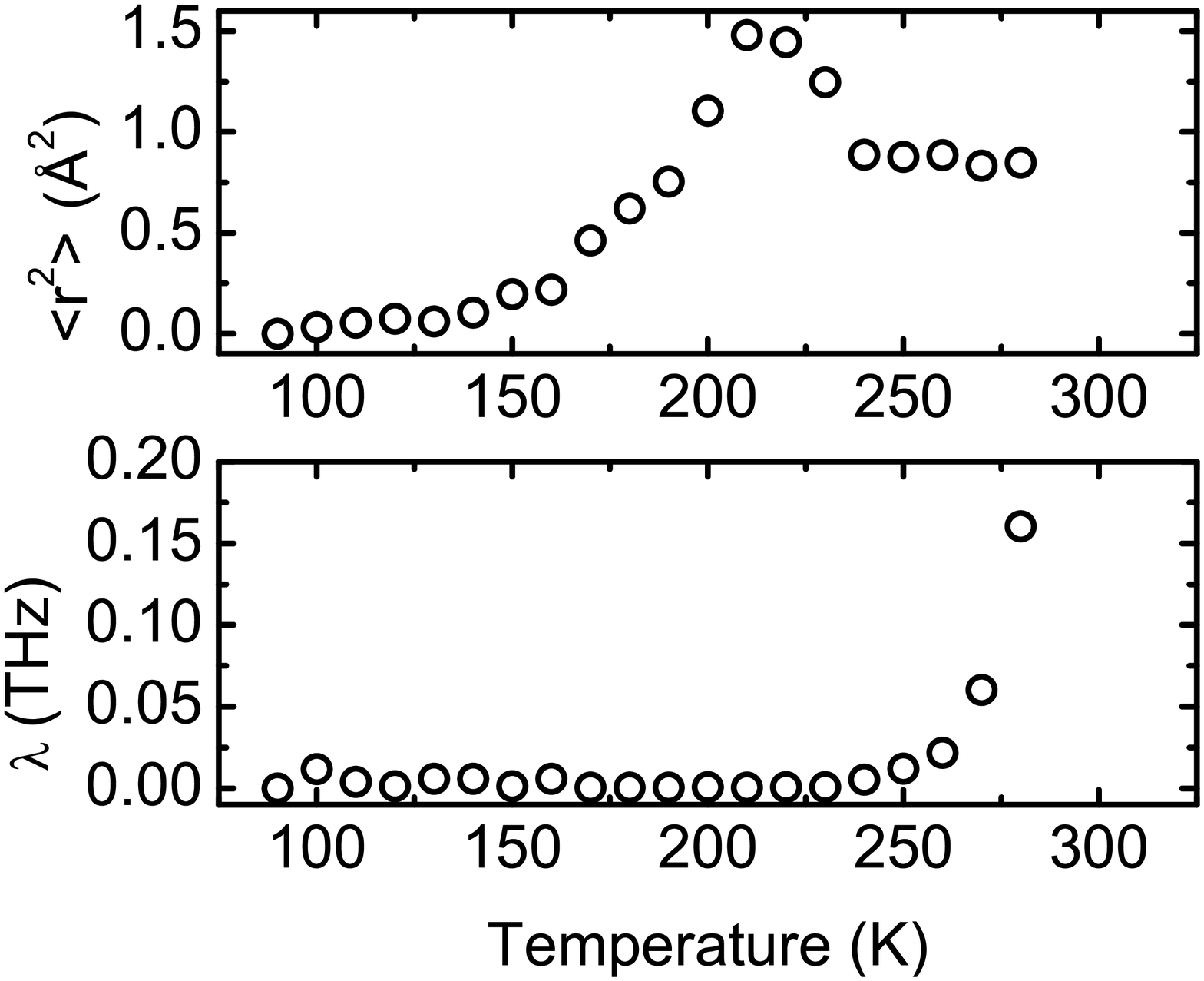}
\vspace{0.8cm}
\caption{MSD, \rsexp, of glutamate dehydrogenase in $CD_3OD/D_2O$ observed by Daniel et. al 
\cite{Daniel:99} on IN16 ($W = 1~\mu$eV) (triangles) and IN6 ($W = 100~\mu$eV) (squares).  Eq.~(\ref{e4}) 
for $\langle r^2\rangle _{slope}$ is fitted to the observed \rsexp~ and the fitted \rsslope~ is shown as a 
solid line. The lower frames show the intrinsic  \rs~ and $\lambda$ obtained from the fit. The extracted 
intrinsic \rs~ decreases above $T$ = 220 K.}
\label{f8}
\end{figure} 

The intrinsic $\langle r^2 \rangle $ shows two interesting features. Firstly, the intrinsic $\langle r^2 
\rangle$ is very similar to the $\langle r^2 \rangle _{exp}$ observed on IN16. The intrinsic $T_D$ is $T_D 
\simeq 150$ K. Secondly, and unexpectedly, the $\langle r^2 \rangle $ is found to decrease for $T \gtrsim 
200$ K and is lower at $T \simeq 250$ K than at $T = 200$ K. This seems unphysical. To explore this 
effect, we arbitrarily kept the intrinsic $\langle r^2 \rangle $ constant at its $200$ K value for 
temperatures $T \gtrsim 200$ K. This $\langle r^2 \rangle$ is shown in Fig.~\ref{f9}. The resulting 
$\langle r^2 \rangle _{slope}$ obtained when $\langle r^2 \rangle$ is held constant for $T > 200$ K is 
shown as the solid line in Fig.~\ref{f9}. The $\langle r^2 \rangle _{slope}$ for $W = 1~\mu$eV (IN16) 
continues to increase above $T = 200$ K but eventually reaches a plateau above $250$ K. This plateau of  
$\langle r^2 \rangle $ at higher temperature appears to be unique for H in glutamate dehydrogenase  or 
there is an issue with the sample or data on IN16 for $T > 200$ K. Further study of this effect would be 
interesting.

\begin{figure}[ht!]
\hspace{0.05cm}
\vspace{-0.5cm}
\includegraphics[scale=0.29,angle=0]{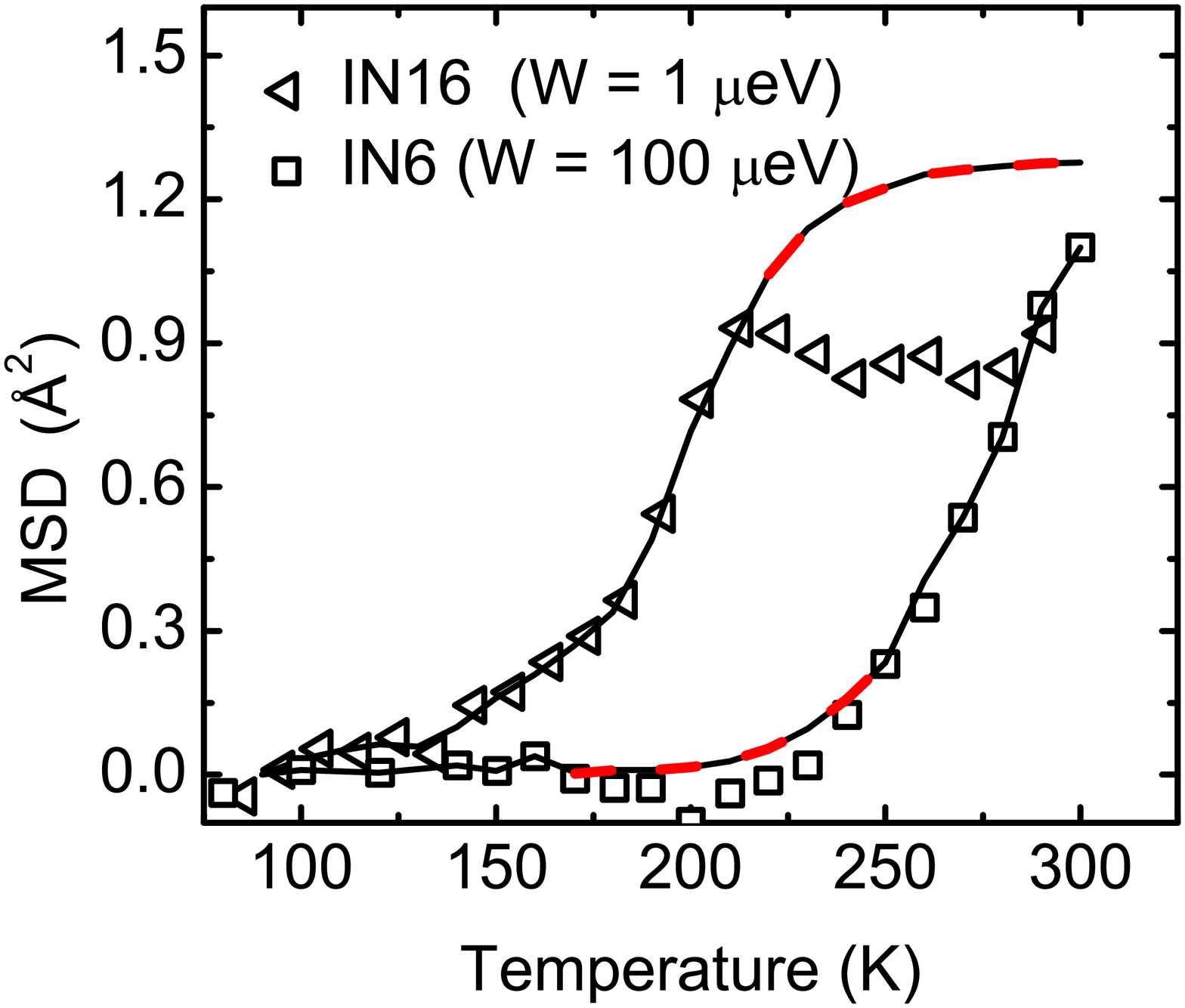}\\
\hspace{0.05cm}
\vspace{-0.5cm}
\includegraphics[scale=0.29,angle=0]{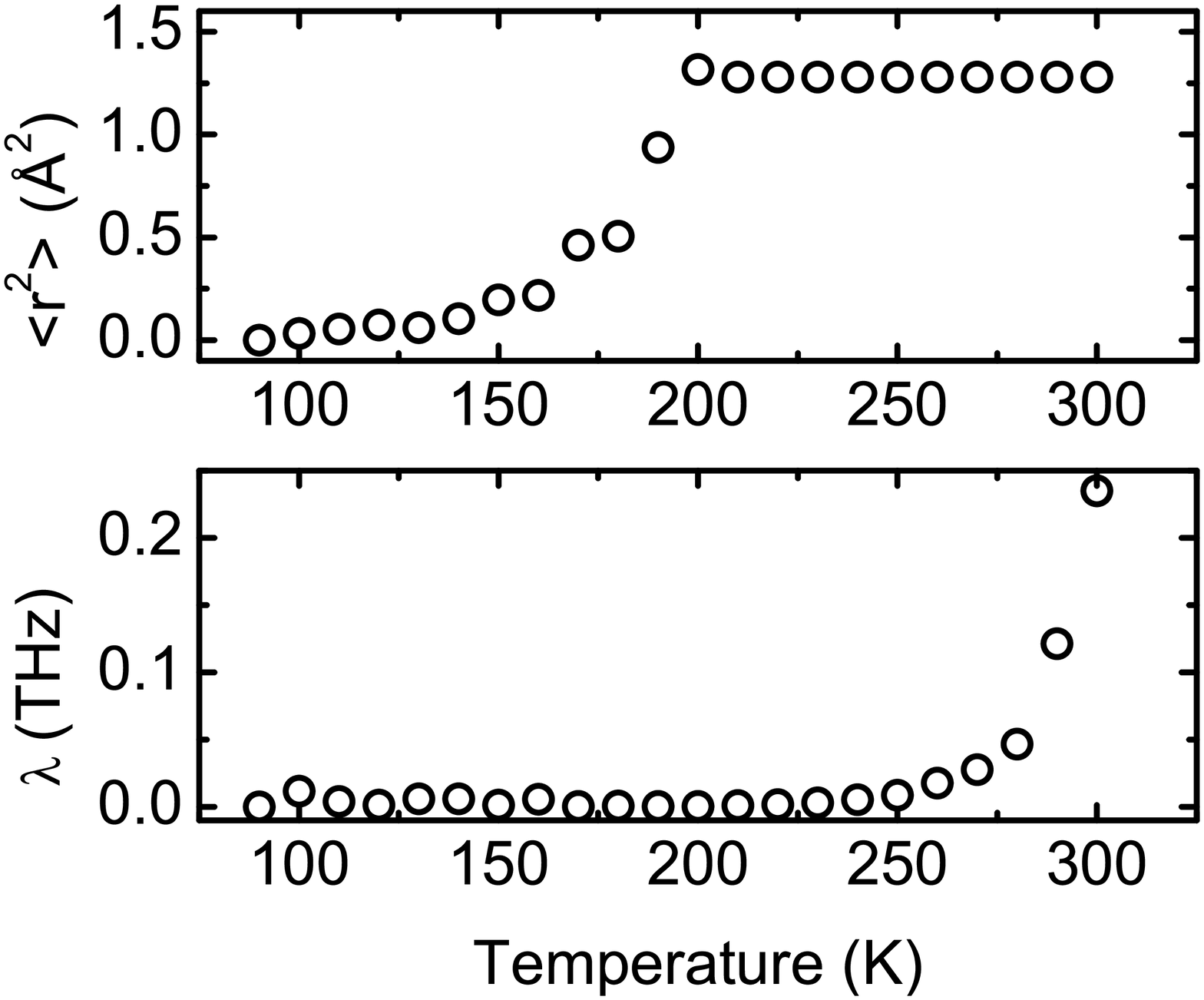}
\vspace{0.8cm}
\caption{The MSD as in Fig.~\ref{f8}. In this case the intrinsic MSD \rs~ is arbitrarily held constant at 
temperatures $T > 200 K$. The solid lines show the values of \rsslope~ obtained for IN16 and IN6 when the 
\rs~ is held constant for $T > 200K$. The expected \rsslope~ continues to increase above 200 K.}
\label{f9}
\end{figure}

The chief result is that an intrinsic $\langle r^2 \rangle$ and $T_D$ independent of instrument resolution 
can be obtained. The intrinsic $T_D$ is the temperature at which the intrinsic $\langle r^2 \rangle $ 
begins to increase markedly. In glutamate dehydrogenase, the intrinsic $T_D$ is at $T_D \simeq 150$ K in 
Fig.~\ref{f8}. This is very close to the $T_D$ observed on IN16 where $W = 1$ $\mu$eV. When $W$ is larger, 
the apparent $T_D$ seen in $\langle r^2 \rangle _{exp}$ is moved to higher temperature. The larger the 
$W$, the higher the apparent $T_D$. This finding is consistent with the results of Becker {\it et 
al.}\cite{Becker:04}. Returning to the previous subsection, we note that the data in Fig.~\ref{f6} shows 
that the larger the $W$, the higher the temperature that is needed to observe a marked increase in 
$\langle r^2 \rangle _{exp}$. That is, the apparent \td~ of \rsexp~ in Fig.~\ref{f6} is shifted to a 
higher temperature for larger W. In this way, the data shown in Figs.~(\ref{f6}) and (\ref{f8}) are 
manifestations of the same phenomena. 

\section{Discussion}

The goal of many measurements of quasielastic neutron scattering from proteins is to determine the thermal 
mean square displacement (MSD) of hydrogen in the protein and in its hydration water. The observed MSD, 
\rsexp, is generally obtained from the slope of the observed elastic incoherent DSF \oqwe~ using 
Eq.~(\ref{e1}). The \oqwe~ is the resolution broadened DSF \sqw~ at $\omega$ = 0. However, the \rsexp~ 
given by Eq.~(\ref{e1}) is equal to the actual \rs~ of the protein only if the scattered intensity exactly 
at $\omega$~= 0 is observed. This is the case only if the energy resolution width, $W$, of the instrument 
is zero so that \oqwe~ reduces to \sqwe. In this limit \oqwe~ is well approximated by its time independent 
part \ii~ given by Eq.~(\ref{e3}), as seen from Eq.~(\ref{e4}), and \rsexp~ equals \rs. When $W$ is finite 
the scattered intensity at finite $\omega$ around $\omega$ = 0 is also observed in \oqwe.  In this case 
the \rsexp~ extracted from Eq.~(\ref{e1}) depends on the resolution width $W$ and is smaller than \rs. 
More precisely, it depends on the ratio of $W$ and to the rate constants that govern the motions in the 
protein. For a given protein, different \rsexp~ are observed on instruments having different resolution 
widths $W$. The impact of a finite $W$ is different in different proteins.

The aim of the present paper is to provide a method in which the intrinsic \rs~ can be extracted from data 
taken on instruments that have different $W$. The essence of the method is to create a model of \oqwe~ in 
which the intrinsic, infinite time \rs~ appears explicitly and to fit the model to the observed \oqwe~ to 
obtain \rs. The model must also contain a description of the motional processes and the resolution width 
$W$. We began with a simple model of the motions and a single, global Debye-Waller factor which led to 
Eq.~(\ref{e2}). By fitting the model to data taken on instruments that have different energy resolutions 
we showed that an instrument independent \rs~ can be obtained. The \rsslope~ given by Eq.~(\ref{e4}) is 
the model equivalent of \rsexp. The model \rsslope~ can also be fitted to observed values of \rsexp~ to 
obtain the intrinsic \rs~ if the measured \oqwe~ are not available. 

\begin{figure}[ht!]
\hspace{0.05cm}
\vspace{-0.5cm}
\includegraphics[scale=0.29,angle=0]{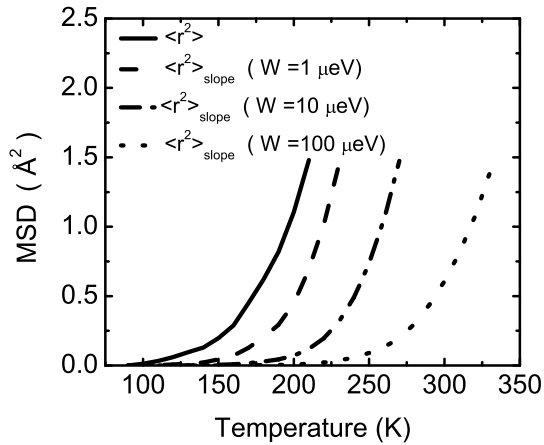}
\vspace{0.8cm}
\caption{The resolution broadened MSD, $\langle r^2\rangle _{slope}$, for different resolution widths, $W$ 
obtained from Eq.(\ref{e4}) illustrating the dependence of the apparent \td~ on W. The intrinsic \rs and 
$\lambda$ of glutamate dehydrogenase, obtained from the fits to data by Daniel et al. \cite{Daniel:99} 
shown in Fig. {\ref{f8}}, are used in Eq.~(\ref{e4}).}
\label{f10}
\end{figure} 

The dependence of the \rsexp~ obtained from the slope of \oqwe~ with $Q^2$ using Eq.~(\ref{e1}), and its 
associated dynamical transition temperature, \td, on the instrument resolution width, $W$ may be 
illustrated and clarified using Eq.~(\ref{e4}) for \rsslope. The \rsslope~ is the model equivalent of  
\rsexp~ given by  Eq.~(\ref{e1}). Firstly, the intrinsic \rs~ is always greater than the observed 
\rsslope~ at a given temperature. The larger is $W$, the further the observed \rsslope~ is suppressed 
below the intrinsic \rs. Similarly, the intrinsic \td~ obtained from the intrinsic \rs~ is always at a 
lower temperature than the observed \td. The larger is $W$, the more the apparent \td~ is shifted to 
higher temperature above the intrinsic \td. The dependence of \rsslope~ and \td~ on $W$ can be illustrated 
by plotting \rsslope~ for increasing values of $W$ using by Eq.~(\ref{e4}) as shown in Fig.~\ref{f10}. As 
input, we use the intrinsic \rs~ of glutamate dehydrogenase shown in Fig.~\ref{f8} plus an extrapolation 
of this \rs~ to higher temperatures. A reasonable $\lambda$ similar to that Fig.~\ref{f8} is also used. 
This intrinsic \rs~ has an intrinsic \td~ $\simeq$ 150 K. From Fig.~\ref{f10}, we see that at a given 
temperature, \rsslope~ calculated from Eq.~(\ref{e4}) always lies below \rs. The larger is $W$, the 
smaller is \rsslope. Equivalently, we could say that \rsslope~ in Fig.~\ref{f10} is shifted to higher 
temperature. The larger the $W$, the further the \rsslope~ is shifted to higher temperature. These shifts 
may be described as a \rsslope~ that lies below \rs~ at a given temperature or a \td~ that is shifted to 
higher temperature than the intrinsic or actual \td. The reduction in \rsslope~ and the apparent increase 
in \td~ with increasing $W$ are one and the same effect.

The impact of a finite $W$ can also be seen in $I(\textbf{Q},t)$ given by Eqs.~ (\ref{e10}), (\ref{e17}) 
and (\ref{e4}). In Eq.~(\ref{e10}), the integral over time is cut off after a time $t_W$~= $W^{-1}$. If 
$W$ is large, $t_W$ is short and long time motions/correlations in $C(t)$ cannot contribute to the 
integral in Eq.~(\ref{e10}). The \oqwe~ then contains only the shorter time motions and \oqwe~ depends on 
$W$. What is important is the ratio of $W$ to $\lambda$. Kneller and Calandrini \cite{Kneller:07} have 
also shown that the magnitude of corrections to the observed $\langle r^2 \rangle _{slope}$ for resolution 
effects depend on the ratio $W/\lambda$. If the intrinsic correlations in the protein (in $C(t)$) decay 
rapidly,  $\lambda~>>~W$, then all correlations in $C(t)$ are included in  Eq.~(\ref{e10}) and the 
integral again becomes independent of $W$. Thus we expect the impact of finite resolution width $W$ to 
vary from protein to protein.

In any model used to extract the intrinsic \rs, some characterization of the rate of decay of correlations 
relative to $W$ is needed. For example, in a White Noise model of correlations, $C(t)$~= $\delta(t)$, 
there are no correlations and any $W$ will capture all correlations and yield the intrinsic \rs. In 
section 3, we saw that the \rsexp~ of HS-0.4 observed on IN16 (W = 1 $\mu$eV) was close to the intrinsic 
\rs. This means that all the correlations in $C(t)$ of HS-0.4 are largely captured within a time scale 
$W^{-1}$ = 4 ps. Whenever the observed \rsexp~ is close to the intrinsic \rs, we may infer that the 
instrument resolution is small enough that all or most of the correlations contributing to $C(t)$ have 
been captured in Eq.~(\ref{e10}).  

The present model is very simple and is intended to be illustrative only. The model has several 
limitations. Firstly, we used a simple representation $C(t) = \exp(-\lambda t)$ to describe the relaxation 
of correlations in the protein. This $C(t)$ is appropriate for a single diffusion process. There is also 
no representation of ballistic propagation or vibration in $C(t)$. The $\lambda$ could also be 
$\textbf{Q}$ dependent. The model could be improved by using a stretched exponential which represents 
several diffusion mechanisms contributing to $C(t)$. However, we found that the data was not significantly 
precise to distinguished between a simple exponential and a stretched exponential $C(t)=\exp (-(\lambda 
t)^{\beta})$. The two $C(t)$ lead to the same intrinsic $ \langle r^2\rangle $. Even better, an 
Ornstein-Uhlenbeck (O-U) function $C(t)=E_{\alpha}(-(\lambda t)^{\alpha})$ could be used. Calandrini et 
al.\cite{Calandrini:08} have found good fits to simulation data and to QENS data using the O-U function 
$E_{\alpha}$ with $\alpha =1/2$. For $\alpha =1/2$, this has an analytic form $E_{1/2}(-(\lambda 
t)^{1/2})=\exp(\lambda t) erfc ((\lambda t)^{1/2})$. We also used this form in Eq. (17) 
($C(t)=\exp(\lambda t) erfc ((\lambda t)^{1/2})$) and found the same value of intrinsic $ \langle 
r^2\rangle $ within precision from fits to data. Current data does not appear to be sufficiently precise 
to distinguish between simple and sophisticated models of $C(t)$ for the purpose of determining $ \langle 
r^2\rangle $. In future applications a more sophisticated $C(t)$ could be incorporated. 

In contrast, we found that a stretched exponential with $\beta = 0.2$ and O-U function with $\alpha =1/2$ 
provide much better fits to simulations than a single exponential. The simulations have much more 
precision especially at larger times. We also found that a stretched exponential with $\beta = 0.2$ and 
$E_{\alpha}$ for $\alpha = 1/2$ are quite similar functions.

The present model uses only a single, global Debye-Waller factor \ii~ in Eq.~(\ref{e3}).  This 
simplification is made by approximating Eq.~(\ref{e6}) by Eq.~(\ref{e7}). In a real protein, the H 
occupies a variety of positions that have different values of \rs. If a distribution of \rs~ values is 
included\cite{Meinhold:08,Kneller:09,Yi:11}, there are terms in \ii~ beyond a simple Gaussian represented 
by a single parameter \rs. In future applications a generalization to include a distribution of \rs~ 
values based on Eq.~(\ref{e6}) should be incorporated.

Improvements to the resolution function used could also be made. Also, recent simulations suggest that 
some motions contribute to \rs~ only after long (picosecond) times. In applications of the present method 
to these proteins, the data from reasonably high resolution instruments may be needed so that information 
on these longer time scale motions is included in the observed \oqwe.      

\section{Summary}

	We have proposed a method for obtaining the intrinsic MSD of H in the proteins independent of the 
instrument resolution width, $W$. The method consists of fitting a model to the observed resolution 
dependent data or resolution dependent MSD. The model contains the intrinsic MSD, the instrument 
resolution width W and a rate constant characterizing the motions of H in the protein. The method is 
applied to existing data in the literature to obtain the intrinsic MSD \rs~ of heparan sulphate (HS-0.4), 
Ribonuclease A and Staphysloccal Nuclase (SNase).

	The intrinsic MSD is defined as the infinite time \rs~ = {$\langle r^2 (t = \infty)\rangle$ that 
appears in the Debye-Waller factor. The intrinsic \rs~ is always greater than the apparent, resolution 
dependent \rsexp. From fits to data at different temperatures, an intrinsic \rs~ as a function of 
temperature is obtained. This intrinsic \rs~ shows a dynamical transition. The intrinsic dynamical 
transition temperature, \td, is defined as temperature at which the intrinsic MSD begins to increase 
markedly with temperature. The intrinsic \td~ is always less than the apparent, resolution dependent \td.

	The essential ingredients of a model needed to obtain \rs~ are (1) a definition of \rs, (2) a 
characteristic time (or times) of the motions that contribute to \rs~ and (3) the instrument resolution 
width. The present model is simple and illustrative only and can be much improved. At the same time, much 
current data in the literature may not be sufficiently precise to distinguish between simple and more 
sophisticated models.

\section{Acknowledgements}

It is a pleasure to acknowledge valuable discussions with Giuseppi Zaccai, Dominique Bicout, Jeremy Smith 
and Edward Lyman. This work was supported by the DOE, Office of Basic Energy Sciences, under contract No. 
ER46680.

\section{Appendix}

The purpose of this appendix is to derive the expression
\begin{equation} \label{A1}
I_{\infty} = \exp (-\frac{1}{3} Q^2\langle r^2\rangle),
\end{equation}
to identify the approximations made to obtain it and to clarify the meaning of \rs. $I_{\infty}$ is the t 
$\rightarrow \infty$ limit of the intermediate incoherent DSF, \iqt~
defined in Eq.~(\ref{e7}). $I(\textbf{Q},\infty)$ is independent of time t. The Fourier transform of \ii~ 
is therefore purely elastic (zero except at $\omega$ = 0) and given by $S_{el}(\textbf{Q},\omega)$ = 
\ii$\delta(\omega)$. This shows that \ii~ arises from structure in the protein since translationally 
invariant systems such as gases and liquids that have no structure have no elastic scattering. Thus the 
motions that contribute to \ii~ and \rs~ are therefore vibrations, hindered rotations, restricted 
diffusion and all motions that in some way are restricted by or reflect a structure. \ii~ is exactly the 
Debye-Waller factor that appears in the scattering of X-rays or neutrons from 
crystals\cite{Debye:12,Waller:28,Maradudin:63}. In crystalline solids, \ii~ is the reduction in intensity 
of a Bragg peak arising from atomic vibration in the solid. For H in proteins, \ii~ is the reduction in 
intensity of the incoherent elastic scattering at any $\textbf{Q}$ value arising from all restricted 
motions of the H.

To obtain \ii, we begin with $I_{\it inc}(\textbf{Q},t)$ given by Eq.~(\ref{e6}). The first approximation 
is to reduce Eq.~(\ref{e6}) to Eq.~(\ref{e7}). If Eq.~(\ref{e6}) were retained then terms beyond the 
Gaussian in Eq.~(\ref{A1}) will be obtained\cite{Meinhold:08,Kneller:09,Yi:11}. Following the arguments 
below Eq.~(\ref{e13}), the t $\rightarrow \infty$ limit of Eq.~(\ref{e7}) is, 
\begin{eqnarray} \label{A2}
I_{\infty} &=& I(\textbf{Q},t = \infty) = \langle \exp(-i\textbf{Q}\cdot\textbf{r}(\infty)) 
\exp(i\textbf{Q}\cdot\textbf{r}(0))\rangle \nonumber \\
   &=& \langle \exp(-i\textbf{Q}\cdot\textbf{r}(0)) \rangle \langle \exp(i\textbf{Q}\cdot\textbf{r}(0))\rangle.
\end{eqnarray}
To arrive at the final line of Eq.~(\ref{A2}) we have assumed: (1) that the position, $\textbf{r}(\infty)$ 
at t $\rightarrow \infty$ of each H is uncorrelated with its position $\textbf{r}(0)$ at $t = 0$ so the 
expectation values  of $\textbf{r}(\infty)$ and $\textbf{r}(0)$ are independent and (2) that the protein 
properties are independent of the time when they are observed so that  {$\langle r^n (\infty) \rangle$} =  
{$\langle r^n (0) \rangle$} = {$\langle r^n \rangle$}. \ii~ is the product of two identical exponentials, 
one containing $i$ and the other $-i$.

To proceed we make a cumulant expansion of each expectation value in \ii,
\begin{equation}\label{A3}
\langle \exp(ix)\rangle = \exp[\sum_{n=1}^{\infty} \frac{(i)^n}{n!} \mu _n] 
\end{equation}
where the $\mu _n$ are cumulants and $x = \pm (\textbf{Q}\cdot\textbf{r})$. The cumulants $\mu _n$ are a 
combination of moments, 
\begin{align}\label{A4}
\nonumber \mu _1 &= \langle x \rangle \\
\nonumber \mu _2 &= \langle x^2 \rangle - \langle x \rangle ^2 \\
\nonumber \mu _3 &= \langle x^3 \rangle - 3\langle x^2 \rangle \langle x \rangle + 2\langle x \rangle ^3 
\\
\mu _4 &= \langle x^4 \rangle - 3\langle x^2 \rangle ^2 - 4 \langle x^3 \rangle \langle x \rangle + 12 
\langle x^2 \rangle \langle x \rangle ^2 - 6\langle x \rangle ^4 .
\end{align}
Because of the $\pm i$ in the exponentials in Eq.~(\ref{A2}), the odd cumulants cancel. In \ii~ only even 
cumulants appear. 
Thus,
\begin{eqnarray}\label{A5}
I_{\infty}&=& \langle \exp(-i\textbf{Q}\cdot\textbf{r}) \rangle \langle \exp(i\textbf{Q}\cdot\textbf{r})\rangle = 
\exp[2\sum_{n~ even}^{\infty} \frac{(i)^n}{n!} \mu _n] \nonumber\\
		  &=& \exp[-\mu _2 + \frac{1}{12} \mu _4 - \frac{2 }{6!} \mu _6 + . . .]
\end{eqnarray} 

We also take $\langle \textbf{r}\rangle$ = 0. Thus up to fourth order,
\begin{equation}\label{A5}
I_{\infty} = \exp[- \langle [\textbf{Q}\cdot\textbf{r}]^2\rangle +\frac{1}{12} [\langle 
[\textbf{Q}\cdot\textbf{r}]^4\rangle - 3\langle [\textbf{Q}\cdot\textbf{r}]^2\rangle ^2 ]
\end{equation}
On neglecting the fourth order cumulant and assuming cubic symmetry so that 
$\langle [\textbf{Q}\cdot\textbf{r}]^2\rangle = Q^2 \langle r_z^2 \rangle = \frac{1}{3} Q^2 \langle 
r^2\rangle$, where the z axis is chosen parallel to $\textbf{Q}$, we obtain \ii~ in Eq.~(\ref{A1}). The 
fourth order cumulant will be small if $\textbf{Q}$ is small or if the motional distribution of the H atom 
is approximately Gaussian. 
%
%

Eq.~(\ref{A1}) defines the intrinsic \rs. This \rs~ is the same as that defined in simulations as the t 
$\rightarrow \infty$ limit of $\frac{1}{2}\langle [\textbf{r}(t) - \textbf{r}(0)]^2\rangle$ since $\langle 
\textbf{r}(t)\cdot\textbf{r}(0)\rangle$ = 0 at 
t $\rightarrow \infty$ and {$\langle r^2 (\infty) \rangle$} =  {$\langle r^2 (0) \rangle$} = {$\langle r^2 
\rangle$}. Explicitly, the intrinsic value of \rs~ is defined as the value that appears in the 
Debye-Waller factor \ii.

\bibliographystyle{apsrev}

\end{document}